\documentstyle[12pt,axodraw]{article}

\begin{document}
\baselineskip 0.7cm

\newcommand{\gsim}{ \mathop{}_{\textstyle \sim}^{\textstyle >} }
\newcommand{\lsim}{ \mathop{}_{\textstyle \sim}^{\textstyle <} }
\newcommand{\bra}{ \langle }
\newcommand{\ket}{ \rangle }
\newcommand{\EV}{ {\rm eV} }
\newcommand{\KEV}{ {\rm keV} }
\newcommand{\MEV}{ {\rm MeV} }
\newcommand{\GEV}{ {\rm GeV} }
\newcommand{\TEV}{ {\rm TeV} }
\renewcommand{\thefootnote}{\fnsymbol{footnote}}
\setcounter{footnote}{1}

\begin{titlepage}
\begin{flushright}
UT-825
\\
September 1998
\end{flushright}

\vskip 0.35cm
\begin{center}
{\large \bf 
Long Lived Superheavy Dark Matter\\ 
with Discrete Gauge Symmetries
}
\vskip 1.2cm
K.~Hamaguchi$^{1}$, Yasunori Nomura$^{1}$, and T.~Yanagida$^{1,2}$
\vskip 0.4cm

$^{1}$ {\it Department of Physics, University of Tokyo,\\
         Tokyo 113-0033, Japan}
\\
$^{2}$ {\it Research Center for the Early Universe, University of Tokyo,\\
         Tokyo 113-0033, Japan}

\vskip 1.5cm

\abstract{The recently observed ultra-high energy (UHE) cosmic rays 
 beyond the Greisen-Zatsepin-Kuzmin bound can be explained by 
 the decays of some superheavy $X$ particles forming a part of dark
 matter in our universe.
 We consider various discrete gauge symmetries ${\bf Z}_N$ to ensure the 
 required long lifetime ($\tau_X \simeq 10^{10}-10^{22}~{\rm years}$) of
 the $X$ particle to explain the UHE cosmic rays in the minimal
 supersymmetric standard model (MSSM) with massive Majorana neutrinos.
 We show that there is no anomaly-free discrete gauge symmetry to make
 the lifetime of the $X$ particle sufficiently long in the MSSM with the
 $X$ particle.
 We find, however, possible solutions to this problem especially 
 by enlarging the particle contents in the MSSM.
 We show a number of solutions introducing an extra pair of singlets
 $Y$ and $\bar{Y}$ which have fractional ${\bf Z}_N$ $(N=2,3)$ charges.
 The present experimental constraints on the $X$ particle are briefly
 discussed.}
\end{center}
\end{titlepage}

\renewcommand{\thefootnote}{\arabic{footnote}}
\setcounter{footnote}{0}

%
%
%
%

\section{Introduction}

The observation of ultra-high energy (UHE) cosmic rays 
\cite{UHE_CR1, UHE_CR2, UHE_CR3} with energies beyond the 
Greisen-Zatsepin-Kuzmin (GZK) bound $E \sim 5 \times 10^{10}~\GEV$
\cite{GZK_cutoff} give a serious challenge to the standard frameworks of
not only astrophysical acceleration mechanisms but also particle physics 
models.
A number of possible solutions to the problem have been considerd 
\cite{UHE_TD, UHE_SH1, UHE_SH2, UHE_M, previous}.
Among them decays of primordial superheavy $X$ particles of mass
$M_X$ ($M_X \sim 10^{13}~\GEV$) \cite{UHE_SH1, UHE_SH2} forming a part
of dark matter in our universe are the most manifest
possibility, although scenarios with topological defects like cosmic
strings may be still viable alternatives \cite{UHE_TD}.

However, the decay scenario of superheavy $X$ particle has obvious
problems.
First of all, the ratio of the present $X$-particle number
density $n_X$ to the entropy $s$ must lie in a range
\cite{UHE_SH2} 
\begin{eqnarray}
  10^{-33} \lsim n_X / s \lsim 10^{-21} 
\end{eqnarray}
to account for the observed UHE cosmic rays 
\cite{UHE_CR1, UHE_CR2, UHE_CR3}.
Although the window for $n_X$ is very wide, a fine tuning is inevitable
to obtain the suitable $n_X$ in the history of the universe.
Recently, this problem has been, however, solved by the authors of
\cite{CKR_production, KT_production}.
They have found that inflation in the early universe may generate a
desirable amount of such a superheavy particle during the reheating
epoch just after the end of inflation, provided 
$M_X \sim H \sim 10^{13}~\GEV$ ($H$ being the Hubble parameter at the
end of inflation).
Since the production mechanism for the superheavy particle involves
only gravitational interactions, it is quite independent of detailed
nature of the particle as well as the reheating process.

The second problem is to explain the required lifetime 
$\tau_X \simeq 10^{10}-10^{22}~{\rm years}$ \cite{UHE_SH2}, which is
abnormally long for such a superheavy particle.
It is, therefore, very natural to impose some symmetries to guarantee
the long lifetime of the superheavy $X$ particle.
In a recent article \cite{previous}, we have shown a model of a
discrete gauge symmetry ${\bf Z}_{10}$ in which a superheavy $X$
particle has naturally a long lifetime 
$\tau_X \simeq 10^{11}-10^{26}~{\rm years}$
for its mass $M_X \simeq 10^{13}-10^{14}~\GEV$.
However, in the previous work \cite{previous}, we have assumed that
neutrinos are all massless.
The purpose of this paper is to extend the previous analysis including
the neutrino mass terms.
We find that there is no anomaly-free discrete gauge symmetry 
${\bf Z}_N$ which explains the required long lifetime of the $X$
particle.
This is because an introduction of the neutrino Majorana masses gives a
stronger constraint on the ${\bf Z}_N$ charges for quarks and leptons in 
the supersymmetric standard model.

We also discuss possible solutions to the above problem.
We show that an introduction of a new pair of singlets $Y$ and $\bar{Y}$
leads to the desired lifetime of the $X$ particle.
We briefly discuss a possible connection of the $X$ particle of mass 
$M_X \sim 10^{13}~\GEV$ to a string inspired particle-physics model.
Throughout this paper we assume supersymmetry (SUSY).

In section 2 we summarize all possible discrete gauge symmetries in the
presence of Majorana mass terms for neutrinos in the minimal SUSY
standard model (MSSM).
We find that only ${\bf Z}_2$ and/or ${\bf Z}_3$ are allowed as
consistent gauge symmetries even if the Green-Schwarz anomaly
cancellation mechanism \cite{GS-PL} works.
In section 3 we introduce one superheavy Majorana-type particle called
$X$ or one pair of Dirac-type particles $X$ and $\bar{X}$.
With a certain assumption for their transformation properties under the 
standard-model gauge group we show that there is no discrete gauge
symmetry giving the desired long lifetime of the $X$ particles of masses 
$M_X \simeq (0.3-3) \times 10^{13}~\GEV$.
In section 4 we discuss possible resolutions of this problem and
find that if we introduce an extra pair of singlets $Y$ and $\bar{Y}$
and assign suitable fractional ${\bf Z}_N$ charges for them, the 
${\bf Z}_3$ gauge symmetry may naturally account for the required long
lifetime of the $X$ particle.
We also show that a product group ${\bf Z}_2 \times {\bf Z}_2$ is a
possible alternative.
The last section 5 is devoted to discussion and conclusions.
We also add an appendix, in which we show the result of our analysis in
the case of neutrinos being massive Dirac fermions.

\section{Possible Discrete Gauge Symmetries in the Supersymmetric
         Standard Model}

In this section we discuss new symmetries accommodated in the MSSM,
which can be regarded as possible candidates that account for naturally
the required long lifetime of a superheavy $X$ particle.
We restrict our consideration to gauge (local) symmetries, since any global 
symmetries are believed to be broken explicitly by topological effects
of gravity \cite{Coleman} and the $X$ particle may no longer survive until
the present as dark matter in our universe.\footnote{
The lifetime of the $X$ particle can be longer
than the age of the universe if topological effects of gravity are
extremely small \cite{UHE_SH1}.}
If these gauge symmetries are spontaneously broken by the
vacuum-expectation value of a field $\phi$, there may be extremely small 
couplings $(\bra \phi \ket/M_*)^n$ 
$(M_* \simeq 10^{18}~\GEV, n \in {\bf Z})$ in the low-energy effective
Lagrangian.
In this case, there are a large number of possibilities realizing the 
desired long lifetime of the $X$ particle.
However, once we admit broken symmetries, our analysis becomes too
complicated and involved.
Thus, we restrict our discussion mainly to unbroken symmetries in this
paper. (In section \ref{long-lived_model} we discuss briefly some broken 
symmetries.)
Then, the symmetries cannot be continuous, 
since no massless gauge boson other than the photon is observed.
This leaves a unique possibility, unbroken discrete gauge symmetries 
\cite{discrete_gauge}.

We now make our analysis on discrete gauge symmetries
in the framework of the MSSM with neutrino mass terms.
As for the neutrino masses, either Dirac or Majorana type is possible.
We mainly consider the Majorana neutrinos, since in the Dirac case we do
not have a natural explanation of the smallness of neutrino masses. 
We give, nevertheless, a discussion in the case of Dirac neutrinos in
the Appendix \ref{Dirac_neutrino_A} for completeness.

The Majorana masses of neutrinos arise from the effective superpotential
\begin{eqnarray}
  W_{\rm eff} = \frac{\kappa_{ij}}{M_R} l_i l_j H_u H_u.
\label{Majorana_mass}
\end{eqnarray}
Here, $l_i, H_u$ are SU(2)$_L$-doublet lepton and Higgs doublet chiral
multiplets, respectively,\footnote{
The vacuum-expectation value of $H_u$, $\bra H_u \ket$, gives rise to
masses for the up-type quarks.
We must introduce the other Higgs doublet $H_d$ to generate masses for
the down-type quarks and the charged leptons.} 
and $i,j = 1,\cdots,3$ are family indices.
$\kappa_{ij}$ denote dimensionless coupling constants and $M_R$ is the
scale at which this operator is generated.

We require the following conditions to possible discrete
gauge symmetries ${\bf Z}_N$ in the MSSM:
\begin{enumerate}
\def\theenumi{\roman{enumi}}
\def\labelenumi{(\theenumi)}
\item All terms present in the MSSM superpotential 
      must be allowed by ${\bf Z}_N$.
\item The Majorana mass terms for neutrinos in 
      eq.~(\ref{Majorana_mass}) are also allowed by ${\bf Z}_N$.
\item All anomalies for ${\bf Z}_N$ should vanish (cancellation with 
      the Green-Schwarz term will be given later).
\end{enumerate}

The first condition (i) reduces the number of independent ${\bf Z}_N$
charges for the MSSM particles as shown in
Table~\ref{discrete_charge}.\footnote{
At first glance, the ${\bf Z}_N$ symmetries seem to be broken by the
vacuum-expectation values of the Higgs doublets resulting in formation
of domain walls.
This is not true, however, since a linear combination of the ${\bf Z}_N$ 
and U(1)$_Y$ remains unbroken, and no domain walls are formed.
That is, we can always choose the ${\bf Z}_N$ charges for $H_u$ and
$H_d$ vanishing by taking a suitable U(1)$_Y$ gauge rotation.}
Here, we have chosen the ${\bf Z}_N$ charge for a first family
SU(2)$_L$-doublet quark $q_1$ to be zero.
This can be always done by using a gauge rotation of U(1)$_Y$ in the
standard model without loss of generality.
Note that the presence of family diagonal and off-diagonal Yukawa
couplings needed in the MSSM suggests the ${\bf Z}_N$
charges for the MSSM particles to be family independent.\footnote{
The mixing among different families are not so clear in the lepton
sector.
The existence of neutrino oscillation \cite{atm_osc, sol_osc}, however,
leads to this conclusion.}
Thus, we omit the family indices, hereater.

The second requirement (ii) restricts the ${\bf Z}_N$ charges for the
MSSM particles as
\begin{eqnarray}
  \left\{
    \begin{array}{@{\,}l} 
      p-m = 0 \qquad\qquad (N : {\rm odd}) \\ \\
      p-m = 0 \;\; {\rm or} \;\; \pm N/2 \qquad\qquad (N : {\rm even}).
    \end{array}
  \right.
\label{Majorana_mass_cond}
\end{eqnarray}
Here, $p, m$ are given in Table~\ref{discrete_charge}.

The third requirement (iii) gives a set of constraints \cite{Ibanez}:
\begin{eqnarray}
  \left\{
    \begin{array}{@{\,}l} 
      0 = \frac{1}{2} r_1 N \\ \\
      -\frac{3}{2}p = \frac{1}{2} r_2 N \qquad\qquad 
        (r_1, r_2, r_3, r_4 \in {\bf Z}) \\ \\
      3(-p+m) = r_3 N + \frac{\eta}{2} r_4 N,
    \end{array}
  \right.
\label{an_cancel_cond}
\end{eqnarray}
where $\eta = 1$ and $0$ for $N = $ even and odd, respectively. 
The first equation comes from the cancellation of $\{ {\bf Z}_N \} 
\{ {\rm SU}(3)_C \}^2$ anomalies, the second from the cancellation of 
$\{ {\bf Z}_N \} \{ {\rm SU}(2)_L \}^2$ anomalies, and the last from the 
cancellation of ${\bf Z}_N$-gravitational anomalies.
Note that the anomaly-free condition for discrete gauge symmetries is
much weaker than that for continuous ones \cite{Ibanez}.

From eqs.~(\ref{Majorana_mass_cond}, \ref{an_cancel_cond}), we find
nontrivial discrete gauge symmetries,\footnote{
We discard a trivial case of $p = m = 0$.}
\begin{eqnarray}
  {\bf Z}_2 &:& (p,m) = (0,1), \label{Z_2}\\
  {\bf Z}_3 &:& (p,m) = (1,1). \label{Z_3}
\end{eqnarray}
There are other solutions which satisfy 
eqs.~(\ref{Majorana_mass_cond}, \ref{an_cancel_cond}), but
they are trivial embeddings of the above ${\bf Z}_2$ or ${\bf Z}_3$ in
higher $N$, or charge conjugation of eq.~(\ref{Z_3}), or direct product
of eq.~(\ref{Z_2}) and eq.~(\ref{Z_3}).
The symmetry (\ref{Z_2}) is nothing but the matter parity, which is
equivalent to the R-parity in the context of the MSSM.
The symmetry (\ref{Z_3}) is the so-called ``baryon parity'' discoverd by 
the authors of \cite{IR-NP} in a different context.
These two symmetries forbid dimension-4 operatrors of baryon-number
violation which cause too rapid proton decay.
Furthermore, the baryon parity forbids proton decay absolutely 
\cite{baryon-parity}.
The discrete charges for the MSSM particles are given in 
Table~\ref{Z2_Z3_charge}.

Now, let us turn to consider the Green-Schwarz term \cite{GS-PL}.
With this term, anomaly cancellation conditions
eq.~(\ref{an_cancel_cond}) change to 
\begin{eqnarray}
  \left\{
    \begin{array}{@{\,}l} 
      0 = \frac{1}{2} r_1 N + k \delta_{GS} \\ \\
      -\frac{3}{2}p = \frac{1}{2} r_2 N + k \delta_{GS} \qquad\qquad 
        (r_1, r_2, r_3, r_4 \in {\bf Z}) \\ \\
      3(-p+m) = r_3 N + \frac{\eta}{2} r_4 N + 24 \delta_{GS},
    \end{array}
  \right.
\label{an_cancel_cond_GS}
\end{eqnarray}
where $k$ is the Kac-Moody level of the SU(3)$_C$ and SU(2)$_L$ gauge
groups, and $\delta_{GS}$ is a constant \cite{Ibanez}.
Here, we have chosen the Kac-Moody levels of SU(3)$_C$ and SU(2)$_L$ to
be the same in view of the gauge coupling unification.
The solutions to these conditions are, however, the same as in the case
without Green-Schwarz term, since non-trivial equation derived from
eqs.~(\ref{Majorana_mass_cond}, \ref{an_cancel_cond_GS}) is equivalent
to that from eqs.~(\ref{Majorana_mass_cond}, \ref{an_cancel_cond}). 
Thus, we conclude that anomaly-free discrete gauge symmetries 
in the MSSM with neutrino mass terms are only the matter parity 
${\bf Z}_2$ and the baryon parity ${\bf Z}_3$.

We finally comment on how the neutrino Majorana mass terms 
eq.~(\ref{Majorana_mass}) are induced.
This can be done by introducing antineutrino chiral multiplets
$\bar{\nu}$ \cite{Seesaw} or SU(2)$_L$-triplet chiral multiplets $\xi$
and $\bar{\xi}$ \cite{triplet-Seesaw, FY_text}.
The discrete charges for these fields are also given in
Table~\ref{Z2_Z3_charge}.
The interactions and masses inducing the operators 
eq.~(\ref{Majorana_mass}) are 
\begin{eqnarray}
  \left\{
    \begin{array}{@{\,}l} 
      W_{\bar{\nu}} = M_R\, \bar{\nu} \bar{\nu} + l \bar{\nu} H_u \\ \\
      W_{\xi} = M_R\, \xi \bar{\xi} + \xi l l + \bar{\xi} H_u H_u,
    \end{array}
  \right.
\end{eqnarray}
respectively.
Then, the neutrino Majorana masses are generated through the diagrams
shown in Fig.~\ref{Majorana_induce}.\footnote{
The Yukawa coupling matrix of $\xi$ may be completely different from
those of the Higgs $H_u$ and $H_d$.
Thus, the observed large mixing in neutrino sector \cite{atm_osc} 
may not be necessarily surprising if the Majorana mass terms for
neutrinos are induced by $\xi$ and $\bar{\xi}$ exchange diagrams.}

\section{Decays of the Superheavy Particles}

In this section we discuss whether we can obtain the required long
lifetime of a superheavy $X$ particle imposing the matter parity 
(${\bf Z}_2$) and/or the baryon parity (${\bf Z}_3$).
We choose the mass of the $X$ particle to be about 
$(0.3-3) \times 10^{13}~\GEV$ in order to explain the UHE cosmic ray by
its decay, using the result in \cite{X-particle}.\footnote{
The mass of the $X$ particle may be somewhat larger than the value in
\cite{X-particle}, since in the present case we consider many-body
decay while in \cite{X-particle} the two-body decay is assumed.}
However, it should be noted that the $M_X$ which gives the best fit for
the UHE cosmic ray spectrum might change depending on the fragmentation
function of quark and gluon jets \cite{fragmentation}.
A detailed analysis including such ambiguity should be made elsewhere.

Given the mass of the $X$ particle, we can calculate the lifetime of the 
$X$ particle once we specify the operators through which the $X$
particle decays into the MSSM particles.
Namely, if the $X$ particle decays through dimension-$n$ operators, 
\begin{eqnarray}
  W = \frac{1}{M_*^{n-4}} X \psi^{n-2} \quad , \quad 
  K = \frac{1}{M_*^{n-4}} X \psi^{n-3},
\label{decay_interaction}
\end{eqnarray}
where $\psi$ denotes the MSSM particles, then the lifetime of the $X$
particle is given by
\begin{eqnarray}
  \tau_X \sim \left( \frac{M_*}{M_X} \right)^{2(n-4)} \frac{1}{M_X}
         \simeq 10^{9n-80}-10^{11n-87}~{\rm years},
\label{decay_lifetime}
\end{eqnarray}
for the cut-off scale $M_* \simeq 10^{18}~\GEV$ and $M_X \simeq
(0.3-3) \times 10^{13}~\GEV$.
Here, $K$ is a K\"{a}hler potential.
Choosing $n = 10$ gives the lifetime 
$\tau_X \simeq 10^{10}-10^{23}~{\rm years}$ while the desired lifetime
is $\tau_X \simeq 10^{10}-10^{22}~{\rm years}$.\footnote{
If we choose the cut-off scale $M_*$ to be the gravitational scale,
${\it i.e.}$ $M_* = 2.4 \times 10^{18}~\GEV$, the lifetime becomes
longer by a factor $10^4$.
However, even if it is the case, the essential point of our conclusion
is not much changed.}
Thus, dimension$=$10 operators are favorable for the $X$ particle 
to be a source of UHE cosmic rays (See Fig.~\ref{lifetime_graph}).

The $X$ particle may be either Majorana- or Dirac- type particle.
We first consider the Majorana case.
In this case, we choose the ${\bf Z}_2$ $({\bf Z}_3)$ charge for the $X$ 
particle to be $0$ or $1$  ($0$ or $3/2$) so that it has the Majorana
mass $(W_X = (M_X/2) X^2)$.
The discrete charge for the $X$ particle should be integer, otherwise
it cannot decay into the MSSM particles.
Thus, we discard the case of the $X$ particle carrying the $3/2$ charge
of ${\bf Z}_3$.
We search for the lowest dimensional operators of the form in 
eq.~(\ref{decay_interaction}) for the symmetries ${\bf Z}_2$, 
${\bf Z}_3$ and ${\bf Z}_2 \times {\bf Z}_3$.
We assume that the $X$ particle transforms as 
$({\bf 8}, {\bf 1})_0$, $({\bf 1}, {\bf 3})_0$ or 
$({\bf 1}, {\bf 1})_0$ under the standard-model gauge group
SU(3)$_C\times$SU(2)$_L\times$U(1)$_Y$.\footnote{
As for the gauge coupling unification see section 5.}
Here, the first and second numbers in each parentheses correspond to
representations of the SU(3)$_C$ and SU(2)$_L$, respectively, and 
the numbers in the subscripts of each parentheses denote the
U(1)$_Y$ hypercharges.
The result is shown in Table~\ref{lowest_operator_Majorana}.
We find that the lifetime of the $X$ particle is always shorter than
$10^{-25}~{\rm sec}$ and the $X$ particle with such a short lifetime
cannot be a source of the UHE cosmic rays.

We consider, now, the Dirac case.
In this case, the ${\bf Z}_2$ $({\bf Z}_3)$ charge for the $X$ particle
is completely free.
(The charges for the $X$ and $\bar{X}$ fields must be opposite to have
an invariant mass.)
If the $X$ particle has fractional charge, however, it cannot decay into 
the MSSM particles.
Thus, we choose ${\bf Z}_2$ $({\bf Z}_3)$ charge for the $X$ particle to 
be integer.
As for the transformation properties under the standard-model gauge
group, we consider that the $X$ and $\bar{X}$ particles form complete
SU(5)$_{\rm GUT}$ multiplets in view of the gauge coupling unification.
Then, in order to maintain the perturbative unification we restrict our
discussion only to the case of the vector-like field $X + \bar{X}$ being 
${\bf 1} + {\bf 1}^*$, ${\bf 5} + {\bf 5}^*$, 
${\bf 10} + {\bf 10}^*$ and ${\bf 15} + {\bf 15}^*$.\footnote{
If one introduces further higher dimensional representations at the
intermediate scale $\sim 10^{13}~\GEV$, gauge coupling constants 
blow up below the grand unification scale $\sim 10^{16}~\GEV$.}
The SU(5)$_{\rm GUT}$ representations ${\bf 1}$, ${\bf 5}$, ${\bf 10}$
and ${\bf 15}$ are decomposed under the standard-model gauge group as
\begin{eqnarray}
  {\bf 1}  &=& ({\bf 1}, {\bf 1})_{0}, \\
  {\bf 5}  &=& ({\bf 3}, {\bf 1})_{-1/3} + ({\bf 1}, {\bf 2})_{1/2}, \\
  {\bf 10} &=& ({\bf 3}, {\bf 2})_{1/6} + ({\bf 3}^*, {\bf 1})_{-2/3} 
               + ({\bf 1}, {\bf 1})_{1}, \\
  {\bf 15} &=& ({\bf 6}, {\bf 1})_{-2/3} + ({\bf 3}, {\bf 2})_{1/6}
               + ({\bf 1}, {\bf 3})_{1}.
\end{eqnarray}
Thus, we search for the operators contributiong to the $X$ particle
decay, provided that the $X$ particle transforms as 
$({\bf 1}, {\bf 1})_{0}$, $({\bf 1}, {\bf 1})_{1}$, 
$({\bf 3}, {\bf 1})_{-1/3}$, $({\bf 3}^*, {\bf 1})_{-2/3}$, 
$({\bf 1}, {\bf 2})_{1/2}$, $({\bf 3}, {\bf 2})_{1/6}$, 
$({\bf 6}, {\bf 1})_{-2/3}$ or $({\bf 1}, {\bf 3})_{1}$ under the
standard-model gauge group and the $\bar{X}$ particle transforms as the
conjugate representation of the $X$.
The lowest dimensional operators are given in
Table~\ref{lowest_operator_Dirac}.
This shows that the $X$ and $\bar{X}$ particles have always lifetimes
shorter than $10^{-3}~{\rm sec}$, and again they cannot be sources of
UHE cosmic rays.

To summarize, we find no anomaly-free discrete gauge symmetry 
${\bf Z}_N$ ensuring the required long lifetime of the $X$ particle to 
explain the UHE cosmic rays in the MSSM with massive Majorana neutrinos.
This is because anomaly-cancellation conditions
(\ref{Majorana_mass_cond}) and (\ref{an_cancel_cond})
are too restrictive to have large $N$ solutions unlike in the previous
work \cite{previous}.
As a consequence the $X$ particle decays into the MSSM particles through 
relatively lower dimensional operators resulting in too short lifetimes.
We discuss possible solutions to this problem in the next section.

\section{Models for the Superheavy Particle with the Desired Lifetime}
\label{long-lived_model}

In the previous section we have found that the unstable $X$ particle
always has a lifetime much shorter than the age of the universe so that
its decay cannot explain the UHE cosmic rays.
Our analysis is based on the following assumptions, however:
\begin{enumerate}
\def\theenumi{\alph{enumi}}
\def\labelenumi{(\theenumi)}
\item The symmetries under consideration are not spontaneously broken.
\item The $X$ particle is an elementary particle up to the cut-off scale 
      $M_* \simeq 10^{18}~\GEV$.
\item There are no extra field other than the MSSM ones and the $X$
      particle.
\end{enumerate}
In this section we show explicitly that if we remove one of the three
assumptions the negative conclusion in the previous section can be
evaded.

At first, we consider the case of broken symmetries.
In this case, symmetries are not necessarily discrete ones, since
gauge bosons associated with the continuous symmetries are massive and
may not be observed.
We can imagine, for example, that there is the so-called fiveness
symmetry\footnote{
A linear combination of this fiveness $n_5$ and the hypercharge $Y$ of
U(1)$_Y$ form $B-L = (4Y-n_5)/5$.}
in the MSSM and it is broken by the vacuum-expectation value of a field
$\phi$ at the scale $M_R$.
The fiveness charges $n_5$ for the MSSM and $\phi$ fields are given in
Table~\ref{fiveness_charge}.
Then, the $X$ particle can couple to the MSSM particles through
operators which may contain the $\phi$ field.
After the condensation of $\phi$ field, the effective operators
contributing to the $X$ particle decay may have small couplings
$(\bra \phi \ket/M_*)^n = (M_R/M_*)^n$ 
$(M_* \simeq 10^{18}~\GEV, n \in {\bf Z})$.
If the $X$ particle transforms as $({\bf 8}, {\bf 1})_0$ under the
standard-model gauge group and has fiveness charge $n_5 = -65$, for
example, the lowest dimensional operators which cause the $X$ particle
decay are 
\begin{eqnarray}
  W_{\rm eff} =   \left\{
    \begin{array}{@{\,}l} 
      \frac{1}{M_*}\left(\frac{\bra \phi \ket}{M_*}\right)^{6}
            Xq\bar{d}l 
            \simeq \frac{10^{-24}}{M_*}Xq\bar{d}l \\ \\
      \frac{1}{M_*}\left(\frac{\bra \phi \ket}{M_*}\right)^{6}
            X\bar{u}\bar{d}\bar{d} 
            \simeq \frac{10^{-24}}{M_*}X\bar{u}\bar{d}\bar{d}
    \end{array}
  \right.
\end{eqnarray}
for $M_R \simeq 10^{14}~\GEV$.
Then, the $X$ particle has the lifetime
\begin{eqnarray}
  \tau_X \sim 10^{48} \left( \frac{M_*}{M_X} \right)^{2} \frac{1}{M_X}
         \simeq 10^{13}-10^{16}~{\rm years},
\end{eqnarray}
and it can be a source of the UHE cosmic rays.
Similar examples may be constructed also in the case of broken discrete
symmetries.
These examples show that if one considers broken symmetries one can
easily explain the UHE cosmic rays by the decay of superheavy
$X$-particle dark matter.
However, this solution may not be very attractive because
of an unnaturally large charge for the $X$ particle.
Thus, we seek for alternative solutions in the remaining part of this
section.

Now we consider another way to generate extremely small coupling
constants in an effective Lagrangian.
Suppose that the $X$ particle is composite particle
rather than elementary one \cite{UHE_M}.
We assume a non-Abelian gauge interaction with matter multiplets $Q$
which transform non-trivially under the gauge group.
Below the dynamical scale $\Lambda$ of the gauge interaction, the
confinement occurs and gauge invariant composite fields become
low-energy degrees of freedom.
We can identify the lightest state among these composite fields with the 
$X$ field.
The operator matching relation is 
\begin{eqnarray}
  X = \left( \frac{4\pi}{\Lambda} \right)^{r-1} Q^{r} \qquad\qquad
      (r \in {\bf Z}),
\end{eqnarray}
provided the $X$ field consists of $r$ $Q$ fields \cite{NDA}.
Thus, if there are allowed tree-level operators
\begin{eqnarray}
  W = \frac{1}{M_*^{r+n-5}} Q^{r} \psi^{n-2} \quad , \quad 
  K = \frac{1}{M_*^{r+n-5}} Q^{r} \psi^{n-3},
\end{eqnarray}
where $\psi$ denotes the MSSM particles, the low-energy effective
operators contributing to the $X$ particle decay are
\begin{eqnarray}
  W_{\rm eff} = \left( \frac{\Lambda}{4\pi M_*} \right)^{r-1} 
                 \frac{1}{M_*^{n-4}} X \psi^{n-2} \quad , \quad 
  K_{\rm eff} = \left( \frac{\Lambda}{4\pi M_*} \right)^{r-1} 
                 \frac{1}{M_*^{n-4}} X \psi^{n-3}.
\end{eqnarray}
The effective coupling constants $( \Lambda/4\pi M_*)^{r-1}$ may be very 
small so that the lifetime of the $X$ particle can be long enough to
explain the UHE cosmic rays even if $n$ is small \cite{UHE_M}.

Finally, we discuss whether we can obtain the required long lifetime of
the $X$ particle keeping the assumptions (a) and (b).
The basic strategy is as follows.
We first choose ${\bf Z}_N$ charge for the $X$ particle so that it
cannot decay into the MSSM particles.
Next, we introduce a pair of vector-like fields $Y$ and $\bar{Y}$
which has a fractional charge of the ${\bf Z}_N$ symmetry, so that the
$X$ particle is able to decay into the MSSM particles through higher
dimensional operators containing $Y$ or $\bar{Y}$ fields, provided that
the mass of these fields is much smaller than that of the $X$ particle.
Then, the lifetime of the $X$ particle may be set long enough to explain
the UHE cosmic rays by choosing suitable ${\bf Z}_N$ charges for the
$Y$ field.
In the following, we show concrete examples of this scenario.
For simplicity, we assume that the $X$ and $Y$ particles transform
as $({\bf 8}, {\bf 1})_0$ and $({\bf 1}, {\bf 1})_0$ under the
standard-model gauge group, respectively.

To begin with, we consider the baryon parity ${\bf Z}_3$ and set its
charge for the $X$ particle to be $3/2$.
Then, the $X$ particle cannot decay into the MSSM particles alone, since 
all MSSM particles have integer ${\bf Z}_3$ charges.
Then, we introduce $Y$ and $\bar{Y}$.
We suppose that they have oposite ${\bf Z}_N$ charge for each other
forming a massive Dirac-type multiplet.
If the $Y$ field has the ${\bf Z}_3$ charge $1/10$, $7/10$, $13/10$ or
$19/10$, for example, the lowest dimensional operator contributing to
the $X$ particle decay is
\begin{eqnarray}
  W = \frac{1}{M_*^6}XYYYYY\bar{u}\bar{d}\bar{d},
\end{eqnarray}
and the lifetime of the $X$ particle is
\begin{eqnarray}
  \tau_X \sim \left( \frac{M_*}{M_X} \right)^{12} \frac{1}{M_X}
         \simeq 10^{10}-10^{23}~{\rm years},
\label{lifetime_with_Y}
\end{eqnarray}
for the cut-off scale $M_* \simeq 10^{18}~\GEV$ and $M_X \simeq
(0.3-3) \times 10^{13}~\GEV$.\footnote{
If we change the cut-off scale from $M_* \simeq 10^{18}~\GEV$, the
dimension of required operators may be also changed from $n=10$ as shown
in eq.~(\ref{decay_lifetime}).
However, the present analysis is also extended to such cases in a
straightforward manner.}
This shows an excellent agreement with the required lifetime of the $X$
particle.
The lowest dimensional operators with a complete list of the ${\bf Z}_3$ 
charge for the $Y$ field are given in
Table~\ref{lowest_operator_Y_Z3},\footnote{
A similar argument is hold also in the case of the $X$ particle
being $({\bf 1}, {\bf 3})_0$.
The operators contributing to the $X$ decay may contain $lH_u$ 
($ll$ or $H_uH_u$) which can be induced by an exchange of the
$\bar{\nu}$ ($\xi$, $\bar{\xi}$) fields.
In this case, the cut-off scale may be rather $M_*^{n-3}M_R$ than
$M_*^{n-4}$ altering the lifetime of the $X$ particle.}
where the $X$ particle has the lifetime given in
eq.~(\ref{lifetime_with_Y}).

The same conclusion can be drawn also by considering a product of two
matter parities ${\bf Z}_2 \times {\bf Z}_2$.
We suppose that the $X$ particle is odd under the first ${\bf Z}_2$ and
even under the second ${\bf Z}_2$ 
($\it{i.e.}$ ${\bf Z}_2(1)\times{\bf Z}_2(0)$) so
that it cannot decay into the MSSM particles without extra fields.
If we introduce the $Y$ and $\bar{Y}$ fields, the lifetime of the $X$
particle can be set as in eq.~(\ref{lifetime_with_Y}) for an appropriate 
choice of the ${\bf Z}_2 \times {\bf Z}_2$ charge for the $Y$ field.
The lowest dimensional operators which cause the $X$ particle decay with 
the desired lifetime in eq.~(\ref{lifetime_with_Y}) are given in
Table~\ref{lowest_operator_Y_Z2_Z2} together with the charges for the
$Y$ fields.

We have shown, in this section, that the decay of the $X$ particle can
explain the UHE cosmic rays when one of the three assumptions (a), (b)
or (c) is removed.
Although all these possibilities require additional structures to the
MSSM, they are not very implausible since we don't know anything about
the physics at the UHE scale.
Thus, we may conclude that there are natural mechanisms to guarantee the 
long lifetime of the superheavy $X$ particle required for explanation of 
the observed UHE cosmic rays.

\section{Discussion and Conclusions}
\label{discussion_conclusion}

The observed ultra-high energy (UHE) cosmic rays 
\cite{UHE_CR1, UHE_CR2, UHE_CR3} beyond the GZK bound seem to be a big
challenge in elementary particle physics.
The simplest explanation is that the UHE cosmic rays originate from
decays of some superheavy $X$ particles forming a part of dark matter in 
our universe \cite{UHE_SH1, UHE_SH2}.
To make this model more fascinating we must account for the long
lifetime of the $X$ particle, 
$\tau_X \simeq 10^{10}-10^{22}~{\rm years}$ \cite{UHE_SH2},
required for a consistent explanation of the observed UHE cosmic rays.
Since the mass of $X$ particle is 
$M_X \simeq (0.3-3) \times 10^{13}~\GEV$ 
\cite{UHE_SH1, UHE_SH2, X-particle, fragmentation}, it is very difficult
to understand the long lifetime of such a superheavy particle without
having symmetry reasons.

Hoping a natural explanation of the long lifetime, we have first
seached for possible discrete gauge symmetries ${\bf Z}_N$ under the
circumstance where neutrinos have tiny Majorana masses suggested from
the recently observed atmospheric and solar neutrino anomalies 
\cite{atm_osc, sol_osc}.
We have shown that there are only two anomaly-free discrete gauge
symmetries, ${\bf Z}_2$ and ${\bf Z}_3$.
We have found, furthermore, no $X$ particle candidate to have the
desired long lifetime and mass within the MSSM.
This is because the discrete gauge symmetries ${\bf Z}_2$ and 
${\bf Z}_3$ are so small that the $X$ particle decays quickly into the
MSSM particles.

We have, however, found possible solutions to the above problem in this
paper.
In particular, we have obtained many solutions by enlarging particle
contents in the MSSM.
We have shown various solutions (in Tables~\ref{lowest_operator_Y_Z3}
and \ref{lowest_operator_Y_Z2_Z2}) introducing an extra pair of singlets
$Y$ and $\bar{Y}$ which have fractional ${\bf Z}_N$ $(N=2,3)$ charges.
In these examples we have assumed that the $X$ particle transforms as 
$({\bf 8}, {\bf 1})_0$ under the standard-model gauge group.
It is remarkable that this $X$ particle raises the gauge coupling
unification scale up to the perturbative string scale 
$\sim 5 \times 10^{17}~\GEV$ together with a partner 
$({\bf 1}, {\bf 3})_0$ \cite{discrepancy}.
We hope that the presence of $X$, $Y$ and $\bar{Y}$ particles and the
discrete gauge symmetry ${\bf Z}_2$ or ${\bf Z}_3$ will be understood by 
some underlying physics at the Planck scale.

Finally, we should comment on experimental constraints on our superheavy 
particle $X$ in the case that dark matter in our universe is dominantly
composed of them.
Then, since the $X$ particle $({\bf 8}, {\bf 1})_{0}$ has strong QCD
interactions, various experiments and cosmological considerations lead
to serious constraints in the parameter space of $M_X$ and $\sigma_p$ 
\cite{SIMP_bound}, with $\sigma_p$ being the scattering cross
section on protons.
In particular, a recent underground experiment has almost excluded the
mass region of our interest ($M_X \simeq (0.3-3) \times 10^{13}~\GEV$)
for the extreme case that the $X$ particles close the universe 
($\Omega_X \simeq 1$) \cite{underground}.
The density $\Omega_X$ of the $X$ particle, for example, must be smaller
than about $10^{-3}$ if $\sigma_p = 10^{-26} {\rm cm^2}$.
Therefore, they may be detected in future experiments if the density of
the $X$ particles $({\bf 8}, {\bf 1})_{0}$ is near its upper bound.

\section*{Acknowledgments}

We would like to thank M.~Teshima for stimulating discussions.
Y.N. thanks the Japan Society for the Promotion of Science for financial 
support.
This work is supported in part by the Grant-in-Aid, Priority Area
``Supersymmetry and Unified Theory of Elementary Particles''(\# 707).

\appendix
\section{The Case of Massive Dirac Neutrinos}
\label{Dirac_neutrino_A}

We consider that the neutrinos are Majorana fermions in the text.
In this appendix we discuss the case of massive Dirac neutrinos and
show that there are some discrete gauge symmetries which explain the 
required long lifetime of the $X$ particles 
$\tau_X \simeq 10^{10}-10^{22}~{\rm years}$.

First, we impose similar anomaly-cancellation conditions as in 
section 2 to discrete gauge symmetries ${\bf Z}_N$. 
Among the conditions (i)-(iii) in section 2, however, the second
conditon (ii) must be replaced by the following new condition:
\begin{enumerate}
\def\theenumi{\roman{enumi}}
\def\labelenumi{(\theenumi)$^{\prime}$}
\item[(ii)$^{\prime}$] The Dirac mass terms for neutrinos 
\begin{eqnarray}
  W = y_{ij}l_i\bar{\nu}_jH_u
\label{Dirac_mass_A}
\end{eqnarray}    
are allowed by ${\bf Z}_N$.
\end{enumerate}
Here $\bar{\nu}_i$ are antineutrino chairal multiplets and $y_{ij}$
denote Yukawa coupling constants, which must be very small
so as to give small neutrino masses.
We omit family indices, hereafter.  

The first two conditions (i) and (ii)$^{\prime}$ reduce the number of
independent ${\bf Z}_N$ charges for the MSSM particles and the
antineutrino chiral multiplets as shown in
Table~\ref{discrete_charge_Dirac_A}.
The third condition (iii) leads to \cite{Ibanez} 
\begin{eqnarray}
  \left\{
    \begin{array}{@{\,}l} 
      0 = \frac{1}{2} r_1 N \\ \\
      -\frac{3}{2}p = \frac{1}{2} r_2 N \qquad\qquad 
        (r_1, r_2, r_3, r_4 \in {\bf Z}) \\ \\
      0 = r_3 N + \frac{\eta}{2} r_4 N.
    \end{array}
  \right.
\label{an_cancel_cond_Dirac_A}
\end{eqnarray}
These conditions come from the cancellations of $\{ {\bf Z}_N \} 
\{ {\rm SU}(3)_C \}^2$, $\{ {\bf Z}_N \} \{ {\rm SU}(2)_L \}^2$ and
${\bf Z}_N$-gravitational anomalies, respectively.

From eq.~(\ref{an_cancel_cond_Dirac_A}), we have many independent
discrete gauge symmetries, 
\begin{eqnarray}
  {\bf Z}_{3n \pm 1} &:& (p,m) = (0,m), \qquad\qquad 
        (m = 1,\cdots,3n \pm 1) \label{sol_1_A} \\
  {\bf Z}_{3n} &:& (p,m) = (0,m),(n,m),(2n,m). \qquad\qquad 
        (m = 1,\cdots,3n) \label{sol_2_A}
\end{eqnarray}
In the Dirac case, there are a larger number of solutions compared with
in the Majorana case, since the condition (ii)$^{\prime}$
is much weaker than (ii).
We should note that the solutions, $p$ and $m$, are not changed even
with the Green-Schwarz term \cite{GS-PL}.\footnote{
In this case,the conditions~(\ref{an_cancel_cond_Dirac_A})
are modified as
\begin{eqnarray}
  \left\{
    \begin{array}{@{\,}l} 
      0 = \frac{1}{2} r_1 N + k \delta_{GS} \\ \\
      -\frac{3}{2}p = \frac{1}{2} r_2 N + k \delta_{GS} \qquad\qquad 
        (r_1, r_2, r_3, r_4 \in {\bf Z}) \\ \\
      0 = r_3 N + \frac{\eta}{2} r_4 N + 24 \delta_{GS}.
    \end{array}
  \right.
\label{an_cancel_cond_GS_A}
\end{eqnarray}
However, solutions to eq.~(\ref{an_cancel_cond_GS_A}) are the same as in
the case without Green-Schwarz term.}
Among various solutions given in eqs.~(\ref{sol_1_A}) and
(\ref{sol_2_A}) we restrict our discussion to the solutions in which the 
antineutrinos $\bar{\nu}$ do not have Majorana masses 
$M_R \bar{\nu} \bar{\nu}$.
This suggests $2(p-m) \neq k N$ $(k \in {\bf Z})$ 
(see Table~\ref{discrete_charge_Dirac_A}).

Now let us turn to the decay of superheavy $X$ particle.
Since we have arbitrarily large $N$ solutions for ${\bf Z}_N$
symmetries, it is easy to find discrete gauge symmetries under
which the $X$ particle has a sufficiently long lifetime.
If we take ${\bf Z}_{10}$ and assume that the $X$ particle is
Majorana-type one which transforms as $({\bf 8}, {\bf 1})_0$
under the standard-model gauge group, for example, the $X$ particle
decays into the MSSM particles including the antineutrinos through
operators given in Table~\ref{lowest_operator_Dirac_A} for 
$(p,m) = (0,1)$ and $(0,3)$.
The lifetime of the $X$ particle is given by 
\begin{eqnarray}
  \tau_X \sim \left( \frac{M_*}{M_X} \right)^{12} \frac{1}{M_X}
         \simeq 10^{10}-10^{23}~{\rm years},
\end{eqnarray}
which shows that the decay of the $X$ particles can produce the UHE
cosmic rays.
Here, we have assumed $M_X \simeq (0.3-3) \times 10^{13}~\GEV$ as in the 
text.

So far, we have put the small Yukawa coupling constants $y_{ij}$ by hand 
to obtain the light neutrinos.
However, there is a mechanism called Dirac-seesaw which generates
extremely small Yukawa coupling constants \cite{FY_text}.
Thus, the case of massive Dirac neutrino may not be very implausible.

\newpage

%
%
%
\newcommand{\Journal}[4]{{\sl #1} {\bf #2} {(#3)} {#4}}
\newcommand{\APJ}{Ap. J.}
\newcommand{\CJP}{Can. J. Phys.}
\newcommand{\NC}{Nuovo Cim.}
\newcommand{\NP}{Nucl. Phys.}
\newcommand{\PL}{Phys. Lett.}
\newcommand{\PR}{Phys. Rev.}
\newcommand{\PRep}{Phys. Rep.}
\newcommand{\PRL}{Phys. Rev. Lett.}
\newcommand{\PTP}{Prog. Theor. Phys.}
\newcommand{\SJNP}{Sov. J. Nucl. Phys.}
\newcommand{\ZP}{Z. Phys.}

%
\newpage
\begin{table}
\begin{center}
\begin{tabular}{|c|ccccccc|}  \hline 
  & $q$ & $\bar{u}$ & $\bar{d}$ & $l$ & $\bar{e}$ & $H_u$ & $H_d$ \\ \hline  
  ${\bf Z}_N$ & $0$ & $-m$ & $m$ & $-p$ & $p+m$ & $m$ & $-m$ \\ \hline
\end{tabular}
\end{center}
\caption{${\bf Z}_N$ discrete charges for the MSSM particles.
 Here, $p,m = 1,\cdots,N-1$.
 $q, \bar{u}, \bar{d}, l$ and $\bar{e}$ denote
 SU(2)$_L$-doublet quark, up-type antiquark, down-type antiquark,
 SU(2)$_L$-doublet lepton and charged antilepton chiral multiplets.
 $H_u$ and $H_d$ are chiral multiplets for Higgs doublets.}
\label{discrete_charge}
\end{table}
\begin{table}
\begin{center}
\begin{tabular}{|c|ccccccc|ccc|}  \hline 
  & $q$ & $\bar{u}$ & $\bar{d}$ & $l$ & $\bar{e}$ & $H_u$ & $H_d$ 
     & $\bar{\nu}$ & $\xi$ & $\bar{\xi}$ \\ \hline  
  ${\bf Z}_2$ & 0 & 1 & 1 & 0 & 1 & 1 & 1 & 1 & 0 & 0 \\ 
  ${\bf Z}_3$ & 0 & 2 & 1 & 2 & 2 & 1 & 2 & 0 & 2 & 1 \\ \hline
\end{tabular}
\end{center}
\caption{Discrete charges for the MSSM particles under the matter 
 parity (${\bf Z}_2$) and the baryon parity (${\bf Z}_3$).
 $q, \bar{u}, \bar{d}, l$ and $\bar{e}$ denote
 SU(2)$_L$-doublet quark, up-type antiquark, down-type antiquark,
 SU(2)$_L$-doublet lepton and charged antilepton chiral multiplets.
 $H_u$ and $H_d$ are chiral multiplets for Higgs doublets. 
 $\bar{\nu}$ is antineutrino chiral multiplet, and $\xi$ and $\bar{\xi}$
 are SU(2)$_L$-triplet chiral multiplets which generate neutrino
 Majorana mass terms.}
\label{Z2_Z3_charge}
\end{table}
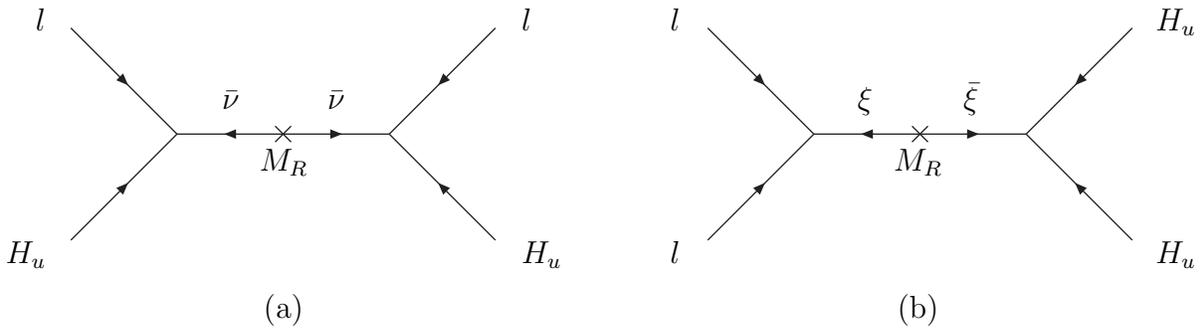
\begin{figure}
\begin{center} 
\begin{picture}(650,120)(-150,120)
  \Text(-60,130)[t]{(a)}
  \ArrowLine(-60,190)(-100,190)
  \Text(-80,200)[b]{$\bar{\nu}$}
  \ArrowLine(-60,190)(-20,190)
  \Text(-40,200)[b]{$\bar{\nu}$}
  \Line(-63,193)(-57,187)
  \Line(-63,187)(-57,193)
  \Text(-60,185)[t]{$M_R$}
  \ArrowLine(-140,230)(-100,190)
  \Text(-150,230)[br]{$l$}
  \ArrowLine(-140,150)(-100,190)
  \Text(-150,150)[tr]{$H_u$}
  \ArrowLine(20,230)(-20,190)
  \Text(30,230)[bl]{$l$}
  \ArrowLine(20,150)(-20,190)
  \Text(30,150)[tl]{$H_u$}
  \Text(180,130)[t]{(b)}
  \ArrowLine(180,190)(140,190)
  \Text(160,200)[b]{$\xi$}
  \ArrowLine(180,190)(220,190)
  \Text(200,200)[b]{$\bar{\xi}$}
  \Line(177,193)(183,187)
  \Line(177,187)(183,193)
  \Text(180,185)[t]{$M_R$}
  \ArrowLine(100,230)(140,190)
  \Text(90,230)[br]{$l$}
  \ArrowLine(100,150)(140,190)
  \Text(90,150)[tr]{$l$}
  \ArrowLine(260,230)(220,190)
  \Text(270,230)[bl]{$H_u$}
  \ArrowLine(260,150)(220,190)
  \Text(270,150)[tl]{$H_u$}
\end{picture}
\caption{The neutrino Majorana mass terms eq.~(\ref{Majorana_mass})can
 be induced by the exchange of (a) anti-neutrino chiral multiplet
 $\bar{\nu}$ or (b) SU(2)$_L$-triplet chiral multiplets $\xi$ and
 $\bar{\xi}$.}
\label{Majorana_induce}
\end{center}
\end{figure}
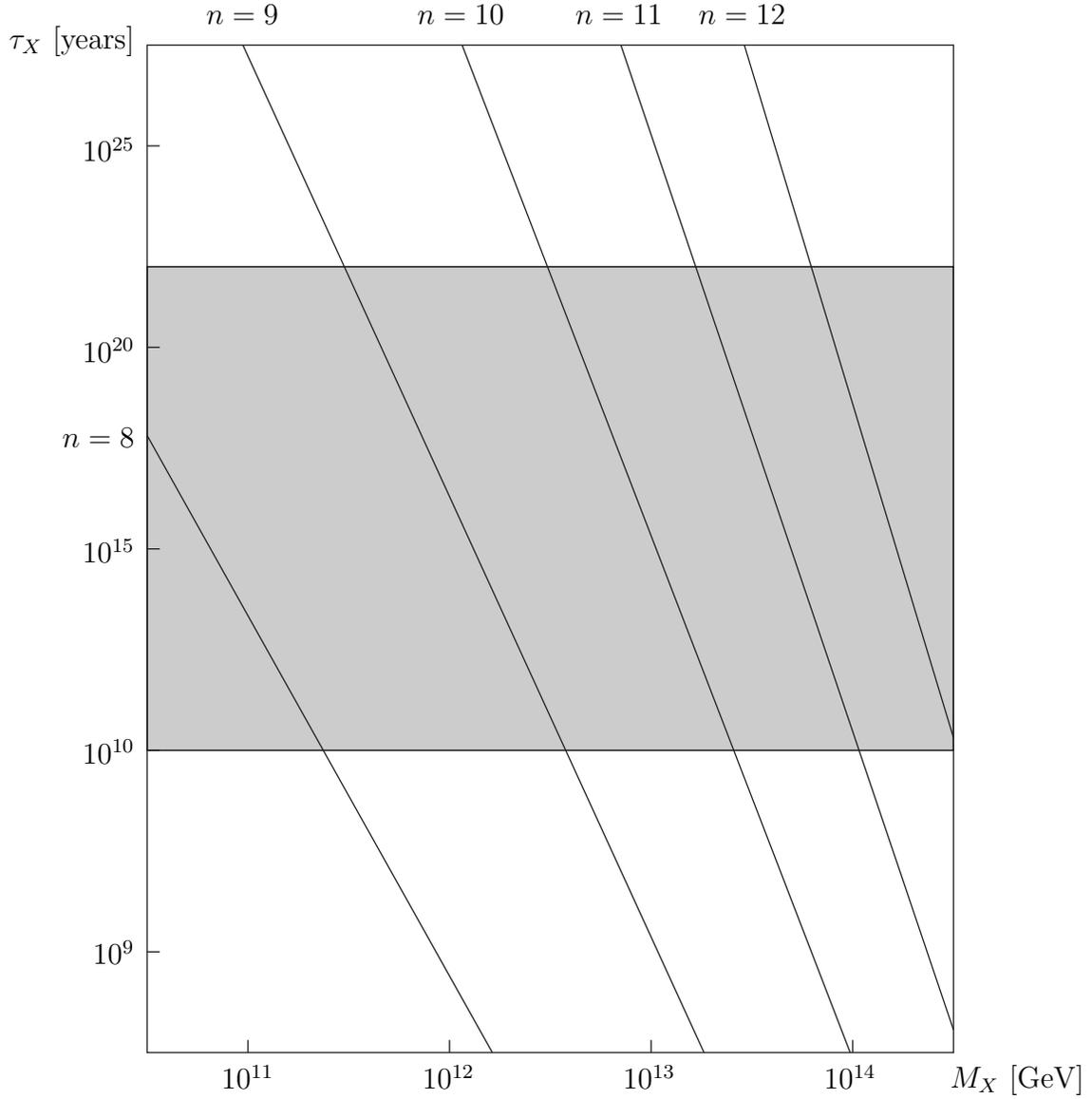
\begin{figure}
\begin{center} 
\begin{picture}(320,500)(0,-20)
  \GBox(0,120)(320,312){0.8}
  \EBox(0,0)(320,400)
  \Text(-5,400)[br]{$\tau_X~[{\rm years}]$} 
  \Text(320,-5)[tl]{$M_X~[\GEV]$}
  \Line(40,0)(40,5)   \Text(40,-5)[t]{$10^{11}$}
  \Line(120,0)(120,5) \Text(120,-5)[t]{$10^{12}$}
  \Line(200,0)(200,5) \Text(200,-5)[t]{$10^{13}$}
  \Line(280,0)(280,5) \Text(280,-5)[t]{$10^{14}$}
  \Line(0,40)(5,40)   \Text(-5,40)[r]{$10^{9}$}
  \Line(0,120)(5,120) \Text(-5,120)[r]{$10^{10}$}
  \Line(0,200)(5,200) \Text(-5,200)[r]{$10^{15}$}
  \Line(0,280)(5,280) \Text(-5,280)[r]{$10^{20}$}
  \Line(0,360)(5,360) \Text(-5,360)[r]{$10^{25}$}
  \Line(0,245)(137,0)
  \Text(-5,245)[r]{$n=8$}
  \Line(38,400)(221,0)
  \Text(38,410)[b]{$n=9$}
  \Line(125,400)(279,0)
  \Text(125,410)[b]{$n=10$}
  \Line(188,400)(320,9)
  \Text(188,410)[b]{$n=11$}
  \Line(237,400)(320,125)
  \Text(237,410)[b]{$n=12$}
\end{picture}
\caption{The lifetime $\tau_X$ of the $X$ particle as a function of its 
 mass $M_X$ in various dimensions $n$ of decay operators 
 ($\tau_X = M_*^{2n-8}/M_X^{2n-7}$).
 Here, we have set $M_* = 10^{18}~\GEV$.
 The shaded region indicates the required lifetime to explain
 the UHE cosmic rays.}
\label{lifetime_graph}
\end{center}
\end{figure}
\begin{table}
\begin{center}
(a) $X({\bf 8}, {\bf 1})_0$ \\
\vspace{0.2cm}
\begin{tabular}{|c|c|c|}  \hline
  symmetries & the lowest dimensional operators & lifetime $\tau_X$ \\ \hline
  ${\bf Z}_2(0)$ & \begin{tabular}{c}
        $K = Xq^{\dagger}q, X\bar{u}^{\dagger}\bar{u}, 
        X\bar{d}^{\dagger}\bar{d}$;\\ 
        $W = Xq\bar{u}H_u, Xq\bar{d}H_d$. \end{tabular} & 
        $10^{-28}-10^{-25}~{\rm sec}$ \\ \hline
  ${\bf Z}_2(1)$ & $W = Xq\bar{d}l, X\bar{u}\bar{d}\bar{d}$. & 
        $10^{-28}-10^{-25}~{\rm sec}$ \\ \hline
  ${\bf Z}_3(0)$ & \begin{tabular}{c}
        $K = Xq^{\dagger}q, X\bar{u}^{\dagger}\bar{u}, 
        X\bar{d}^{\dagger}\bar{d}$;\\
        $W = Xq\bar{u}H_u, Xq\bar{d}l, Xq\bar{d}H_d$. \end{tabular} & 
        $10^{-28}-10^{-25}~{\rm sec}$ \\ \hline
  ${\bf Z}_2(0)\times{\bf Z}_3(0)$ & \begin{tabular}{c}
        $K = Xq^{\dagger}q, X\bar{u}^{\dagger}\bar{u}, 
        X\bar{d}^{\dagger}\bar{d}$;\\ 
        $W = Xq\bar{u}H_u, Xq\bar{d}H_d$. \end{tabular} & 
        $10^{-28}-10^{-25}~{\rm sec}$ \\ \hline
  ${\bf Z}_2(1)\times{\bf Z}_3(0)$ & $W = Xq\bar{d}l$. & 
        $10^{-28}-10^{-25}~{\rm sec}$ \\ \hline
\end{tabular}
\\
\vspace{1cm}
(b) $X({\bf 1}, {\bf 3})_0$ \\
\vspace{0.2cm}
\begin{tabular}{|c|c|c|}  \hline
  symmetries & the lowest dimensional operators & lifetime $\tau_X$ \\ \hline
  ${\bf Z}_2(0)$ & $W = XH_uH_d$. & 
        $10^{-37}-10^{-36}~{\rm sec}$ \\ \hline
  ${\bf Z}_2(1)$ & $W = XlH_u$. & 
        $10^{-37}-10^{-36}~{\rm sec}$ \\ \hline
  ${\bf Z}_3(0)$ & $W = XlH_u, XH_uH_d$. & 
        $10^{-37}-10^{-36}~{\rm sec}$ \\ \hline
  ${\bf Z}_2(0)\times{\bf Z}_3(0)$ & $W = XH_uH_d$. & 
        $10^{-37}-10^{-36}~{\rm sec}$ \\ \hline
  ${\bf Z}_2(1)\times{\bf Z}_3(0)$ & $W = XlH_u$. & 
        $10^{-37}-10^{-36}~{\rm sec}$ \\ \hline
\end{tabular}
\\
\vspace{1cm}
(c) $X({\bf 1}, {\bf 1})_0$ \\
\vspace{0.2cm}
\begin{tabular}{|c|c|c|}  \hline
  symmetries & the lowest dimensional operators & lifetime $\tau_X$ \\ \hline
  ${\bf Z}_2(0)$ & $W = XH_uH_d$. & 
        $10^{-37}-10^{-36}~{\rm sec}$ \\ \hline
  ${\bf Z}_2(1)$ & $W = XlH_u$. & 
        $10^{-37}-10^{-36}~{\rm sec}$ \\ \hline
  ${\bf Z}_3(0)$ & $W = XlH_u, XH_uH_d$. & 
        $10^{-37}-10^{-36}~{\rm sec}$ \\ \hline
  ${\bf Z}_2(0)\times{\bf Z}_3(0)$ & $W = XH_uH_d$. & 
        $10^{-37}-10^{-36}~{\rm sec}$ \\ \hline
  ${\bf Z}_2(1)\times{\bf Z}_3(0)$ & $W = XlH_u$. & 
        $10^{-37}-10^{-36}~{\rm sec}$ \\ \hline
\end{tabular}
\end{center}
\caption{The lowest dimensional operators through which the $X$ particle 
 decays into the MSSM particles under the various discrete gauge
 symmetries.
 The $X$ particle transforms under the standard-model gauge group as 
 (a) $({\bf 8}, {\bf 1})_0$, (b) $({\bf 1}, {\bf 3})_0$ or 
 (c) $({\bf 1}, {\bf 1})_0$.
 The numbers $n_X$ in each parentheses ${\bf Z}_N(n_X)$ denote ${\bf Z}_N$ 
 charges for the $X$ particle.
 Here, we have set the cut-off scale $M_* = 1$ in the table.}
\label{lowest_operator_Majorana}
\end{table}
\begin{table}
\begin{center}
(a) $X({\bf 1}, {\bf 1})_{0} \quad [\; \oplus \; 
    \bar{X}({\bf 1}, {\bf 1})_{0} \;]$\\
\vspace{0.1cm}
\begin{tabular}{|c|c|c|}  \hline
  symmetries & the lowest dimensional operators & lifetime $\tau_X$ \\ \hline
  ${\bf Z}_2(0)$ & $W = XH_uH_d, \bar{X}H_uH_d$. &
        $10^{-37}-10^{-36}~{\rm sec}$ \\ \hline
  ${\bf Z}_2(1)$ & $W = XlH_u, \bar{X}lH_u$. &
        $10^{-37}-10^{-36}~{\rm sec}$ \\ \hline
  ${\bf Z}_3(0)$ & $W = XlH_u, XH_uH_d, \bar{X}lH_u, \bar{X}H_uH_d$. &
        $10^{-37}-10^{-36}~{\rm sec}$ \\ \hline
  ${\bf Z}_3(1)$ & $W = \bar{X}\bar{u}\bar{d}\bar{d}$. &
        $10^{-28}-10^{-25}~{\rm sec}$ \\ \hline
  ${\bf Z}_3(2)$ & $W = X\bar{u}\bar{d}\bar{d}$. &
        $10^{-28}-10^{-25}~{\rm sec}$ \\ \hline
  ${\bf Z}_2(0)\times{\bf Z}_3(0)$ & $W = XH_uH_d, \bar{X}H_uH_d$. &
        $10^{-37}-10^{-36}~{\rm sec}$ \\ \hline
  ${\bf Z}_2(0)\times{\bf Z}_3(1)$ & $W = Xqqql, 
        \bar{X}\bar{u}\bar{u}\bar{d}\bar{e}$. &
        $10^{-19}-10^{-14}~{\rm sec}$ \\ \hline
  ${\bf Z}_2(0)\times{\bf Z}_3(2)$ & $W = X\bar{u}\bar{u}\bar{d}\bar{e}, 
        \bar{X}qqql$. &
        $10^{-19}-10^{-14}~{\rm sec}$ \\ \hline
  ${\bf Z}_2(1)\times{\bf Z}_3(0)$ & $W = XlH_u, \bar{X}lH_u$. &
        $10^{-37}-10^{-36}~{\rm sec}$ \\ \hline
  ${\bf Z}_2(1)\times{\bf Z}_3(1)$ & $W = \bar{X}\bar{u}\bar{d}\bar{d}$. &
        $10^{-28}-10^{-25}~{\rm sec}$ \\ \hline
  ${\bf Z}_2(1)\times{\bf Z}_3(2)$ & $W = X\bar{u}\bar{d}\bar{d}$. &
        $10^{-28}-10^{-25}~{\rm sec}$ \\ \hline
\end{tabular}
\\
\vspace{0.5cm}
(b) $X({\bf 1}, {\bf 1})_{1} \quad [\; \oplus \; 
     \bar{X}({\bf 1}, {\bf 1})_{-1} \;]$ \\
\vspace{0.1cm}
\begin{tabular}{|c|c|c|}  \hline
  symmetries & the lowest dimensional operators & lifetime $\tau_X$ \\ \hline
  ${\bf Z}_2(0)$ & $W = Xll$. &
        $10^{-37}-10^{-36}~{\rm sec}$ \\ \hline
  ${\bf Z}_2(1)$ & $W = \bar{X}\bar{e}$. &
        $\star\quad\star\quad\star$ \\ \hline
  ${\bf Z}_3(0)$ & $W = \bar{X}\bar{d}\bar{d}\bar{d}$. &
        $10^{-28}-10^{-25}~{\rm sec}$ \\ \hline
  ${\bf Z}_3(1)$ & $W = X\bar{u}\bar{u}\bar{d}$. &
        $10^{-28}-10^{-25}~{\rm sec}$ \\ \hline
  ${\bf Z}_3(2)$ & $W = \bar{X}\bar{e}$. &
        $\star\quad\star\quad\star$ \\ \hline
  ${\bf Z}_2(0)\times{\bf Z}_3(0)$ & 
        $W = \bar{X}\bar{u}\bar{d}\bar{d}\bar{e}$. &
        $10^{-19}-10^{-14}~{\rm sec}$ \\ \hline
  ${\bf Z}_2(0)\times{\bf Z}_3(1)$ & 
        $W = X\bar{u}\bar{u}\bar{u}\bar{e}$. &
        $10^{-19}-10^{-14}~{\rm sec}$ \\ \hline
  ${\bf Z}_2(0)\times{\bf Z}_3(2)$ & $W = Xll$. &
        $10^{-37}-10^{-36}~{\rm sec}$ \\ \hline
  ${\bf Z}_2(1)\times{\bf Z}_3(0)$ & 
        $W = \bar{X}\bar{d}\bar{d}\bar{d}$. &
        $10^{-28}-10^{-25}~{\rm sec}$ \\ \hline
  ${\bf Z}_2(1)\times{\bf Z}_3(1)$ & 
        $W = X\bar{u}\bar{u}\bar{d}$. &
        $10^{-28}-10^{-25}~{\rm sec}$ \\ \hline
  ${\bf Z}_2(1)\times{\bf Z}_3(2)$ & $W = \bar{X}\bar{e}$. &
        $\star\quad\star\quad\star$ \\ \hline
\end{tabular}
\\
\vspace{0.5cm}
(c) $X({\bf 3}, {\bf 1})_{-1/3} \quad [\; \oplus \; 
     \bar{X}({\bf 3}^*, {\bf 1})_{1/3} \;]$ \\
\vspace{0.1cm}
\begin{tabular}{|c|c|c|}  \hline
  symmetries & the lowest dimensional operators & lifetime $\tau_X$ \\ \hline
  ${\bf Z}_2(0)$ & $W = Xqq, X\bar{u}\bar{e}, 
        \bar{X}ql, \bar{X}\bar{u}\bar{d}$. &
        $10^{-37}-10^{-36}~{\rm sec}$ \\ \hline
  ${\bf Z}_2(1)$ & $W = X\bar{d}$. &
        $\star\quad\star\quad\star$ \\ \hline
  ${\bf Z}_3(0)$ & $W = Xqq, \bar{X}\bar{u}\bar{d}$. &
        $10^{-37}-10^{-36}~{\rm sec}$ \\ \hline
  ${\bf Z}_3(1)$ & $W = X\bar{u}\bar{d}\bar{d}\bar{d}$. &
        $10^{-19}-10^{-14}~{\rm sec}$ \\ \hline
  ${\bf Z}_3(2)$ & $W = X\bar{d}$. &
        $\star\quad\star\quad\star$ \\ \hline
  ${\bf Z}_2(0)\times{\bf Z}_3(0)$ & $W = Xqq, 
        \bar{X}\bar{u}\bar{d}$. &
        $10^{-37}-10^{-36}~{\rm sec}$ \\ \hline
  ${\bf Z}_2(0)\times{\bf Z}_3(1)$ & 
        $W = X\bar{u}\bar{d}\bar{d}\bar{d}$. &
        $10^{-19}-10^{-14}~{\rm sec}$ \\ \hline
  ${\bf Z}_2(0)\times{\bf Z}_3(2)$ & $W = X\bar{u}\bar{e}, 
        \bar{X}ql$. &
        $10^{-37}-10^{-36}~{\rm sec}$ \\ \hline
  ${\bf Z}_2(1)\times{\bf Z}_3(0)$ & 
        $W = \bar{X}\bar{u}\bar{u}\bar{e}$. &
        $10^{-28}-10^{-25}~{\rm sec}$ \\ \hline
  ${\bf Z}_2(1)\times{\bf Z}_3(1)$ & 
        $W = X\bar{u}\bar{u}\bar{d}\bar{d}\bar{e}, 
        \bar{X}\bar{u}\bar{u}\bar{d}\bar{d}\bar{d}$. &
        $10^{-10}-10^{-3}~{\rm sec}$ \\ \hline
  ${\bf Z}_2(1)\times{\bf Z}_3(2)$ & $W = X\bar{d}$. &
        $\star\quad\star\quad\star$ \\ \hline
\end{tabular}
\end{center}
\end{table}
\begin{table}
\begin{center}
(d) $X({\bf 3}^*, {\bf 1})_{-2/3} \quad [\; \oplus \; 
    \bar{X}({\bf 3}, {\bf 1})_{2/3} \;]$ \\
\vspace{0.1cm}
\begin{tabular}{|c|c|c|}  \hline
  symmetries & the lowest dimensional operators & lifetime $\tau_X$ \\ \hline
  ${\bf Z}_2(0)$ & $W = X\bar{d}\bar{d}$. &
        $10^{-37}-10^{-36}~{\rm sec}$ \\ \hline
  ${\bf Z}_2(1)$ & $W = \bar{X}\bar{u}$. &
        $\star\quad\star\quad\star$ \\ \hline
  ${\bf Z}_3(0)$ & $W = Xqqqq, 
        \bar{X}\bar{u}\bar{u}\bar{d}\bar{d}$. &
        $10^{-19}-10^{-14}~{\rm sec}$ \\ \hline
  ${\bf Z}_3(1)$ & $W = X\bar{d}\bar{d}$. &
        $10^{-37}-10^{-36}~{\rm sec}$ \\ \hline
  ${\bf Z}_3(2)$ & $W = \bar{X}\bar{u}$. &
        $\star\quad\star\quad\star$ \\ \hline
  ${\bf Z}_2(0)\times{\bf Z}_3(0)$ & $W = Xqqqq, 
        \bar{X}\bar{u}\bar{u}\bar{d}\bar{d}$. &
        $10^{-19}-10^{-14}~{\rm sec}$ \\ \hline
  ${\bf Z}_2(0)\times{\bf Z}_3(1)$ & $W = X\bar{d}\bar{d}$. &
        $10^{-37}-10^{-36}~{\rm sec}$ \\ \hline
  ${\bf Z}_2(0)\times{\bf Z}_3(2)$ & \begin{tabular}{c}
        $K = Xql^{\dagger}, X\bar{d}^{\dagger}\bar{e}, 
        \bar{X}q^{\dagger}l, \bar{X}\bar{d}\bar{e}^{\dagger}$;\\
        $W = Xq\bar{e}H_d, \bar{X}\bar{u}lH_u, 
        \bar{X}\bar{d}lH_d$. \end{tabular} &
        $10^{-28}-10^{-25}~{\rm sec}$ \\ \hline
  ${\bf Z}_2(1)\times{\bf Z}_3(0)$ & 
        $W = X\bar{u}\bar{d}\bar{d}\bar{d}\bar{d}, 
        \bar{X}\bar{u}\bar{u}\bar{u}\bar{d}\bar{e}$. &
        $10^{-10}-10^{-3}~{\rm sec}$ \\ \hline
  ${\bf Z}_2(1)\times{\bf Z}_3(1)$ & $W = X\bar{u}\bar{d}\bar{e}$. &
        $10^{-28}-10^{-25}~{\rm sec}$ \\ \hline
  ${\bf Z}_2(1)\times{\bf Z}_3(2)$ & $W = \bar{X}\bar{u}$. &
        $\star\quad\star\quad\star$ \\ \hline
\end{tabular}
\\
\vspace{2.0cm}
(e) $X({\bf 1}, {\bf 2})_{1/2} \quad [\; \oplus \; 
    \bar{X}({\bf 1}, {\bf 2})_{-1/2} \;]$ \\
\vspace{0.1cm}
\begin{tabular}{|c|c|c|}  \hline
  symmetries & the lowest dimensional operators & lifetime $\tau_X$ \\ \hline
  ${\bf Z}_2(0)$ & $W = Xl$. &
        $\star\quad\star\quad\star$ \\ \hline
  ${\bf Z}_2(1)$ & $W = XH_d, \bar{X}H_u$. &
        $\star\quad\star\quad\star$ \\ \hline
  ${\bf Z}_3(0)$ & $W = \bar{X}qqq$. &
        $10^{-28}-10^{-25}~{\rm sec}$ \\ \hline
  ${\bf Z}_3(1)$ & $W = Xl, XH_d, \bar{X}H_u$. &
        $\star\quad\star\quad\star$ \\ \hline
  ${\bf Z}_3(2)$ & \begin{tabular}{c}
        $K = Xq\bar{d}^{\dagger}\bar{d}^{\dagger}, 
        \bar{X}q^{\dagger}\bar{d}\bar{d}$;\\
        $W = \bar{X}\bar{u}\bar{d}\bar{d}H_u, 
        \bar{X}\bar{d}\bar{d}\bar{d}l, 
        \bar{X}\bar{d}\bar{d}\bar{d}H_d$. \end{tabular} &
        $10^{-19}-10^{-14}~{\rm sec}$ \\ \hline
  ${\bf Z}_2(0)\times{\bf Z}_3(0)$ & $W = \bar{X}qqq$. &
        $10^{-28}-10^{-25}~{\rm sec}$ \\ \hline
  ${\bf Z}_2(0)\times{\bf Z}_3(1)$ & $W = Xl$. &
        $\star\quad\star\quad\star$ \\ \hline
  ${\bf Z}_2(0)\times{\bf Z}_3(2)$ & \begin{tabular}{c}
        $K = Xq\bar{d}^{\dagger}\bar{d}^{\dagger}, 
        \bar{X}q^{\dagger}\bar{d}\bar{d}$;\\
        $W = \bar{X}\bar{u}\bar{d}\bar{d}H_u, 
        \bar{X}\bar{d}\bar{d}\bar{d}H_d$. \end{tabular} &
        $10^{-19}-10^{-14}~{\rm sec}$ \\ \hline
  ${\bf Z}_2(1)\times{\bf Z}_3(0)$ & $W = X\bar{u}\bar{d}\bar{d}l$. &
        $10^{-19}-10^{-14}~{\rm sec}$ \\ \hline
  ${\bf Z}_2(1)\times{\bf Z}_3(1)$ & $W = XH_d, \bar{X}H_u$. &
        $\star\quad\star\quad\star$ \\ \hline
  ${\bf Z}_2(1)\times{\bf Z}_3(2)$ & 
        $W = \bar{X}\bar{d}\bar{d}\bar{d}l$. &
        $10^{-19}-10^{-14}~{\rm sec}$ \\ \hline
\end{tabular}
\end{center}
\end{table}
\begin{table}
\begin{center}
(f) $X({\bf 3}, {\bf 2})_{1/6} \quad [\; \oplus \; 
    \bar{X}({\bf 3}^*, {\bf 2})_{-1/6} \;]$ \\
\vspace{0.1cm}
\begin{tabular}{|c|c|c|}  \hline
  symmetries & the lowest dimensional operators & lifetime $\tau_X$ \\ \hline
  ${\bf Z}_2(0)$ & $W = \bar{X}q$. &
        $\star\quad\star\quad\star$ \\ \hline
  ${\bf Z}_2(1)$ & $W = X\bar{d}l$. &
        $10^{-37}-10^{-36}~{\rm sec}$ \\ \hline
  ${\bf Z}_3(0)$ & $W = \bar{X}q$. &
        $\star\quad\star\quad\star$ \\ \hline
  ${\bf Z}_3(1)$ & \begin{tabular}{c}
        $K = Xq\bar{d}^{\dagger}, \bar{X}q^{\dagger}\bar{d}$;\\
        $W = Xqql, XqqH_d, \bar{X}\bar{u}\bar{d}H_u,$\\ 
        $\bar{X}\bar{d}\bar{d}l, \bar{X}\bar{d}\bar{d}H_d$. 
        \end{tabular} &
        $10^{-28}-10^{-25}~{\rm sec}$ \\ \hline
  ${\bf Z}_3(2)$ & \begin{tabular}{c}
        $K = Xq^{\dagger}q^{\dagger}q^{\dagger}\bar{d}, 
        Xq^{\dagger}\bar{u}\bar{d}\bar{d},$\\ 
        $\bar{X}qqq\bar{d}^{\dagger}, 
        \bar{X}q\bar{u}^{\dagger}\bar{d}^{\dagger}\bar{d}^{\dagger}$;\\
        $W = X\bar{u}\bar{u}\bar{d}\bar{d}H_u, 
        X\bar{u}\bar{d}\bar{d}\bar{d}l, 
        X\bar{u}\bar{d}\bar{d}\bar{d}H_d,$\\ 
        $\bar{X}qqqql, \bar{X}qqqqH_d$. \end{tabular} &
        $10^{-10}-10^{-3}~{\rm sec}$ \\ \hline
  ${\bf Z}_2(0)\times{\bf Z}_3(0)$ & $W = \bar{X}q$. &
        $\star\quad\star\quad\star$ \\ \hline
  ${\bf Z}_2(0)\times{\bf Z}_3(1)$ & $W = Xqql, 
        \bar{X}\bar{d}\bar{d}l$. &
        $10^{-28}-10^{-25}~{\rm sec}$ \\ \hline
  ${\bf Z}_2(0)\times{\bf Z}_3(2)$ & 
        $W = X\bar{u}\bar{d}\bar{d}\bar{d}l, \bar{X}qqqql$. &
        $10^{-10}-10^{-3}~{\rm sec}$ \\ \hline
  ${\bf Z}_2(1)\times{\bf Z}_3(0)$ & $W = X\bar{d}l$. &
        $10^{-37}-10^{-36}~{\rm sec}$ \\ \hline
  ${\bf Z}_2(1)\times{\bf Z}_3(1)$ & \begin{tabular}{c}
        $K = Xq\bar{d}^{\dagger}, \bar{X}q^{\dagger}\bar{d}$;\\
        $W = XqqH_d, \bar{X}\bar{u}\bar{d}H_u, 
        \bar{X}\bar{d}\bar{d}H_d$. \end{tabular} &
        $10^{-28}-10^{-25}~{\rm sec}$ \\ \hline
  ${\bf Z}_2(1)\times{\bf Z}_3(2)$ & \begin{tabular}{c}
        $K = Xq^{\dagger}q^{\dagger}q^{\dagger}\bar{d}, 
        Xq^{\dagger}\bar{u}\bar{d}\bar{d},$\\
        $\bar{X}qqq\bar{d}^{\dagger}, 
        \bar{X}q\bar{u}^{\dagger}\bar{d}^{\dagger}\bar{d}^{\dagger}$;\\
        $W = X\bar{u}\bar{u}\bar{d}\bar{d}H_u, 
        X\bar{u}\bar{d}\bar{d}\bar{d}H_d, 
        \bar{X}qqqqH_d$. \end{tabular} &
        $10^{-10}-10^{-3}~{\rm sec}$ \\ \hline
\end{tabular}
\\
\vspace{1cm}
(g) $X({\bf 6}, {\bf 1})_{-2/3} \quad [\; \oplus \; 
    \bar{X}({\bf 6}^*, {\bf 1})_{2/3} \;]$ \\
\vspace{0.1cm}
\begin{tabular}{|c|c|c|}  \hline
  symmetries & the lowest dimensional operators & lifetime $\tau_X$ \\ \hline
  ${\bf Z}_2(0)$ & $W = X\bar{d}\bar{d}$. &
        $10^{-37}-10^{-36}~{\rm sec}$ \\ \hline
  ${\bf Z}_2(1)$ & $W = Xqq\bar{d}, X\bar{u}\bar{d}\bar{e}$. &
        $10^{-28}-10^{-25}~{\rm sec}$ \\ \hline
  ${\bf Z}_3(0)$ & $W = Xqqqq, \bar{X}\bar{u}\bar{u}\bar{d}\bar{d}$. &
        $10^{-19}-10^{-14}~{\rm sec}$ \\ \hline
  ${\bf Z}_3(1)$ & $W = X\bar{d}\bar{d}$. &
        $10^{-37}-10^{-36}~{\rm sec}$ \\ \hline
  ${\bf Z}_3(2)$ & $W = Xqq\bar{d}$. &
        $10^{-28}-10^{-25}~{\rm sec}$ \\ \hline
  ${\bf Z}_2(0)\times{\bf Z}_3(0)$ & 
        $W = Xqqqq, \bar{X}\bar{u}\bar{u}\bar{d}\bar{d}$. & 
        $10^{-19}-10^{-14}~{\rm sec}$ \\ \hline
  ${\bf Z}_2(0)\times{\bf Z}_3(1)$ & $W = X\bar{d}\bar{d}$. &
        $10^{-37}-10^{-36}~{\rm sec}$ \\ \hline
  ${\bf Z}_2(0)\times{\bf Z}_3(2)$ & $W = Xqq\bar{u}\bar{e}, 
        \bar{X}q\bar{u}\bar{d}l$. &
        $10^{-19}-10^{-14}~{\rm sec}$ \\ \hline
  ${\bf Z}_2(1)\times{\bf Z}_3(0)$ & 
        $W = X\bar{u}\bar{d}\bar{d}\bar{d}\bar{d}, 
        \bar{X}\bar{u}\bar{u}\bar{u}\bar{d}\bar{e}$. &
        $10^{-10}-10^{-3}~{\rm sec}$ \\ \hline
  ${\bf Z}_2(1)\times{\bf Z}_3(1)$ & $W = X\bar{u}\bar{d}\bar{e}$. &
        $10^{-28}-10^{-25}~{\rm sec}$ \\ \hline
  ${\bf Z}_2(1)\times{\bf Z}_3(2)$ & $W = Xqq\bar{d}$. &
        $10^{-28}-10^{-25}~{\rm sec}$ \\ \hline
\end{tabular}
\end{center}
\end{table}
\begin{table}
\begin{center}
(h) $X({\bf 1}, {\bf 3})_{1} \quad [\; \oplus \; 
    \bar{X}({\bf 1}, {\bf 3})_{-1} \;]$ \\
\vspace{0.2cm}
\begin{tabular}{|c|c|c|}  \hline
  symmetries & the lowest dimensional operators & lifetime $\tau_X$ \\ \hline
  ${\bf Z}_2(0)$ & $W = Xll, XH_dH_d, \bar{X}H_uH_u$. &
        $10^{-37}-10^{-36}~{\rm sec}$ \\ \hline
  ${\bf Z}_2(1)$ & $W = XlH_d$. &
        $10^{-37}-10^{-36}~{\rm sec}$ \\ \hline
  ${\bf Z}_3(0)$ & \begin{tabular}{c}
        $K = Xqq\bar{d}^{\dagger}\bar{e}^{\dagger}, 
        Xq\bar{d}^{\dagger}\bar{d}^{\dagger}l,$\\ 
        $Xq\bar{d}^{\dagger}\bar{d}^{\dagger}H_u^{\dagger}, 
        Xq\bar{d}^{\dagger}\bar{d}^{\dagger}H_d,$\\
        $\bar{X}q^{\dagger}q^{\dagger}\bar{d}\bar{e}, 
        \bar{X}q^{\dagger}\bar{d}\bar{d}l^{\dagger},$\\ 
        $\bar{X}q^{\dagger}\bar{d}\bar{d}H_u, 
        \bar{X}q^{\dagger}\bar{d}\bar{d}H_d^{\dagger}$;\\
        $W = \bar{X}\bar{u}\bar{d}\bar{d}H_uH_u, 
        \bar{X}\bar{d}\bar{d}\bar{d}lH_u, 
        \bar{X}\bar{d}\bar{d}\bar{d}H_uH_d$. \end{tabular} &
        $10^{-10}-10^{-3}~{\rm sec}$ \\ \hline
  ${\bf Z}_3(1)$ & \begin{tabular}{c}
        $K = Xq^{\dagger}q^{\dagger}\bar{u}, 
        \bar{X}qq\bar{u}^{\dagger}$;\\
        $W = \bar{X}qqqH_u$. \end{tabular} &
        $10^{-19}-10^{-14}~{\rm sec}$ \\ \hline
  ${\bf Z}_3(2)$ & $W = Xll, XlH_u, XH_dH_d, \bar{X}H_uH_u$. &
        $10^{-37}-10^{-36}~{\rm sec}$ \\ \hline
  ${\bf Z}_2(0)\times{\bf Z}_3(0)$ & \begin{tabular}{c}
        $K = Xqq\bar{d}^{\dagger}\bar{e}^{\dagger}, 
        Xq\bar{d}^{\dagger}\bar{d}^{\dagger}l,$\\ 
        $\bar{X}q^{\dagger}q^{\dagger}\bar{d}\bar{e}, 
        \bar{X}q^{\dagger}\bar{d}\bar{d}l^{\dagger}$;\\
        $W = \bar{X}\bar{d}\bar{d}\bar{d}lH_u$. \end{tabular} &
        $10^{-10}-10^{-3}~{\rm sec}$ \\ \hline
  ${\bf Z}_2(0)\times{\bf Z}_3(1)$ & \begin{tabular}{c}
        $K = Xq^{\dagger}q^{\dagger}q^{\dagger}l, 
        Xq^{\dagger}q^{\dagger}\bar{d}\bar{e}^{\dagger},$\\ 
        $Xq^{\dagger}\bar{u}\bar{u}l^{\dagger}, 
        Xq^{\dagger}\bar{u}\bar{d}l,$\\
        $\bar{X}qqql^{\dagger}, 
        \bar{X}qq\bar{d}^{\dagger}\bar{e},$\\ 
        $\bar{X}q\bar{u}^{\dagger}\bar{u}^{\dagger}l, 
        \bar{X}q\bar{u}^{\dagger}\bar{d}^{\dagger}l^{\dagger}$;\\
        $W = X\bar{u}\bar{u}\bar{d}lH_u, X\bar{u}\bar{d}\bar{d}lH_d, 
        \bar{X}qqq\bar{e}H_d$. \end{tabular} &
        $10^{-10}-10^{-3}~{\rm sec}$ \\ \hline
  ${\bf Z}_2(0)\times{\bf Z}_3(2)$ & 
        $W = Xll, XH_dH_d, \bar{X}H_uH_u$. &
        $10^{-37}-10^{-36}~{\rm sec}$ \\ \hline
  ${\bf Z}_2(1)\times{\bf Z}_3(0)$ & \begin{tabular}{c}
        $K = Xq\bar{d}^{\dagger}\bar{d}^{\dagger}H_u^{\dagger}, 
        Xq\bar{d}^{\dagger}\bar{d}^{\dagger}H_d,$\\
        $\bar{X}q^{\dagger}\bar{d}\bar{d}H_u, 
        \bar{X}q^{\dagger}\bar{d}\bar{d}H_d^{\dagger}$;\\
        $W = \bar{X}\bar{u}\bar{d}\bar{d}H_uH_u, 
        \bar{X}\bar{d}\bar{d}\bar{d}H_uH_d$. \end{tabular} &
        $10^{-10}-10^{-3}~{\rm sec}$ \\ \hline
  ${\bf Z}_2(1)\times{\bf Z}_3(1)$ & \begin{tabular}{c}
        $K = Xq^{\dagger}q^{\dagger}\bar{u}, 
        \bar{X}qq\bar{u}^{\dagger}$;\\
        $W = \bar{X}qqqH_u$. \end{tabular} &
        $10^{-19}-10^{-14}~{\rm sec}$ \\ \hline
  ${\bf Z}_2(1)\times{\bf Z}_3(2)$ & $W = XlH_d$. &
        $10^{-37}-10^{-36}~{\rm sec}$ \\ \hline
\end{tabular}
\end{center}
\caption{The lowest dimensional operators through which the $X$
 ($\bar{X}$) particle decays into the MSSM particles under the various
 discrete gauge symmetries.
 The $X$ particle transforms under the standard-model gauge group as
 (a) $({\bf 1}, {\bf 1})_{0}$, (b) $({\bf 1}, {\bf 1})_{1}$, 
 (c) $({\bf 3}, {\bf 1})_{-1/3}$, (d) $({\bf 3}^*, {\bf 1})_{-2/3}$, 
 (e) $({\bf 1}, {\bf 2})_{1/2}$, (f) $({\bf 3}, {\bf 2})_{1/6}$, 
 (g) $({\bf 6}, {\bf 1})_{-2/3}$ or (h) $({\bf 1}, {\bf 3})_{1}$.
 The numbers $n_X$ of each parentheses ${\bf Z}_N(n_X)$ denote ${\bf Z}_N$
 charges for the $X$ particle and the charges for the $\bar{X}$ are
 $-n_X$.
 Here, we have set the cut-off scale $M_* = 1$ in the table.}
\label{lowest_operator_Dirac}
\end{table}
\begin{table}
\begin{center}
\begin{tabular}{|c|ccccccc|c|}  \hline 
  & $q$ & $\bar{u}$ & $\bar{d}$ & $l$ & $\bar{e}$ & $H_u$ & $H_d$ &
        $\phi$ \\ \hline  
  fiveness & $-1$ & $-1$ & $3$ & $3$ & $-1$ & $2$ & $-2$ & $10$ \\ \hline
\end{tabular}
\end{center}
\caption{Fiveness charges $n_5$ for the MSSM particles.
 $q, \bar{u}, \bar{d}, l$ and $\bar{e}$ denote
 SU(2)$_L$-doublet quark, up-type antiquark, down-type antiquark,
 SU(2)$_L$-doublet lepton and charged antilepton chiral multiplets.
 $H_u$ and $H_d$ are chiral multiplets for Higgs doublets.
 $\phi$ breaks the fiveness symmetry by its vacuum-expectation value
 $\bra \phi \ket = M_R$.
 The $B-L$ is given by $(4Y-n_5)/5$.}
\label{fiveness_charge}
\end{table}
\begin{table}
\begin{center}
${\bf Z}_3(3/2) \quad 
    [\; X({\bf 8}, {\bf 1})_{0} \;]$ \\
\vspace{0.1cm}
\begin{tabular}{|c|c|}  \hline
  ${\bf Z}_3$ charges for $Y$ &
        the lowest dimensional operators \\ \hline
  $\frac{1}{10}, \frac{7}{10}, \frac{13}{10}, \frac{19}{10}$ &
        $W = XY^5\bar{u}\bar{d}\bar{d}$. \\ \hline
  $\frac{3}{10}, \frac{9}{10}, \frac{21}{10}, \frac{27}{10}$ &
        \begin{tabular}{c}
        $K = XY^5q^{\dagger}q, 
        XY^5\bar{u}^{\dagger}\bar{u}, 
        XY^5\bar{d}^{\dagger}\bar{d}$, \\
        $XY^{\dagger 5}q^{\dagger}q, 
        XY^{\dagger 5}\bar{u}^{\dagger}\bar{u}, 
        XY^{\dagger 5}\bar{d}^{\dagger}\bar{d}$, \\
        $X\bar{Y}^5q^{\dagger}q, 
        X\bar{Y}^5\bar{u}^{\dagger}\bar{u}, 
        X\bar{Y}^5\bar{d}^{\dagger}\bar{d}$, \\
        $X\bar{Y}^{\dagger 5}q^{\dagger}q, 
        X\bar{Y}^{\dagger 5}\bar{u}^{\dagger}\bar{u}, 
        X\bar{Y}^{\dagger 5}\bar{d}^{\dagger}\bar{d}$; \\
        $W = XY^5q\bar{u}H_u, XY^5q\bar{d}l, XY^5q\bar{d}H_d$,\\
        $X\bar{Y}^5q\bar{u}H_u, X\bar{Y}^5q\bar{d}l, 
        X\bar{Y}^5q\bar{d}H_d$.
        \end{tabular} \\ \hline
  $\frac{11}{10}, \frac{17}{10}, \frac{23}{10}, \frac{29}{10}$ &
        $W = X\bar{Y}^5\bar{u}\bar{d}\bar{d}$. \\ \hline
\end{tabular} 
\end{center}
\caption{The lowest dimensional operators through which the $X$
 particle decays into the MSSM and the $Y$ particles under 
 the ${\bf Z}_3(3/2)$.
 (The number $n_X$ in parenthesis ${\bf Z}_3(n_X)$ denotes ${\bf Z}_3$ 
 charge for the $X$ particle.)
 The $X$ particle transforms under the standard-model gauge group as
 $({\bf 8}, {\bf 1})_{0}$ and the $Y$ and $\bar{Y}$ are singlets 
 $({\bf 1}, {\bf 1})_{0}$.
 For each case, the lifetime of the $X$ particle $\tau_X$ is 
 $10^{10}-10^{23}~{\rm years}$.
 Here, we have set the cut-off scale $M_* = 1$ in the table.}
\label{lowest_operator_Y_Z3}
\end{table}
\begin{table}
\begin{center}
${\bf Z}_2(1) \times {\bf Z}_2(0) \quad 
    [\; X({\bf 8}, {\bf 1})_{0} \;]$ \\
\vspace{0.1cm}
\begin{tabular}{|c|c|}  \hline
  $({\bf Z}_2, {\bf Z}_2)$ charges for $Y$ &
        the lowest dimensional operators \\ \hline
  $(\frac{2n_1}{5}, \frac{2n_2+1}{5})$ &
        \begin{tabular}{c}
        $W = XY^5q\bar{d}\l, XY^5\bar{u}\bar{d}\bar{d}$, \\
        $X\bar{Y}^5q\bar{d}\l, X\bar{Y}^5\bar{u}\bar{d}\bar{d}$.
        \end{tabular} \\ \hline
  $(\frac{2n_2+1}{5}, \frac{2n_1}{5})$ &
        \begin{tabular}{c}
        $K = XY^5q^{\dagger}q, 
        XY^5\bar{u}^{\dagger}\bar{u}, 
        XY^5\bar{d}^{\dagger}\bar{d}$, \\
        $XY^{\dagger 5}q^{\dagger}q, 
        XY^{\dagger 5}\bar{u}^{\dagger}\bar{u}, 
        XY^{\dagger 5}\bar{d}^{\dagger}\bar{d}$, \\
        $X\bar{Y}^5q^{\dagger}q, 
        X\bar{Y}^5\bar{u}^{\dagger}\bar{u}, 
        X\bar{Y}^5\bar{d}^{\dagger}\bar{d}$, \\
        $X\bar{Y}^{\dagger 5}q^{\dagger}q, 
        X\bar{Y}^{\dagger 5}\bar{u}^{\dagger}\bar{u}, 
        X\bar{Y}^{\dagger 5}\bar{d}^{\dagger}\bar{d}$; \\
        $W = XY^5q\bar{u}H_u, XY^5q\bar{d}H_d$,\\
        $X\bar{Y}^5q\bar{u}H_u, X\bar{Y}^5q\bar{d}H_d$.
        \end{tabular} \\ \hline
\end{tabular} 
\end{center}
\caption{The lowest dimensional operators through which the $X$
 particle decays into the MSSM and the $Y$ particles 
 under the ${\bf Z}_2(1) \times {\bf Z}_2(0)$.
 (The numbers $n_X$ in parentheses ${\bf Z}_2(n_X)$ denote ${\bf Z}_2$ 
 charges for the $X$ particle.)
 The $X$ particle transforms under the standard-model gauge group as
 $({\bf 8}, {\bf 1})_{0}$ and the $Y$ and $\bar{Y}$ are singlets 
 $({\bf 1}, {\bf 1})_{0}$.
 Here, $n_1, n_2 = 0,\cdots,4$ except for $(n_1,n_2) = (0,2)$.
 For each case, the lifetime of the $X$ particle $\tau_X$ is 
 $10^{10}-10^{23}~{\rm years}$.
 Here, we have set the cut-off scale $M_* = 1$ in the table.}
\label{lowest_operator_Y_Z2_Z2}
\end{table}
\begin{table}
\begin{center}
\begin{tabular}{|c|cccccccc|}  \hline 
  & $q$ & $\bar{u}$ & $\bar{d}$ & $l$ & $\bar{e}$ & $\bar{\nu}$ & $H_u$ 
  & $H_d$ \\ \hline ${\bf Z}_N$ & $0$ & $-m$ & $m$ & $-p$ & $p+m$ & $p-m$ 
  & $m$ & $-m$ \\ \hline
\end{tabular}
\end{center}
\caption{${\bf Z}_N$ discrete charges for the MSSM particles and 
 the antineutrino.
 Here, $p,m = 1,\cdots,N-1$.
 $q, \bar{u}, \bar{d}, l, \bar{e}$ and $\bar{\nu}$ denote
 SU(2)$_L$-doublet quark, up-type antiquark, down-type antiquark,
 SU(2)$_L$-doublet lepton, charged antilepton and antineutrino 
 chiral multiplets.
 $H_u$ and $H_d$ are chiral multiplets for Higgs doublets.}
\label{discrete_charge_Dirac_A}
\end{table}
\begin{table}
\begin{center}
${\bf Z}_{10}(5) \quad [ X({\bf 8}, {\bf 1})_0 ]$ \\
\vspace{0.1cm}
\begin{tabular}{|c|c|}  \hline
  $(p,m)$ & the lowest dimensional operators \\ \hline
  \begin{tabular}{c} $(0,1)$, \\ $(0,3)$ \end{tabular} & 
  \begin{tabular}{c}
        $K = Xq^{\dagger}q\bar{\nu}\bar{\nu}\bar{\nu}\bar{\nu}\bar{\nu},
        Xq^{\dagger}q\bar{\nu}^{\dagger}\bar{\nu}^{\dagger}
        \bar{\nu}^{\dagger}\bar{\nu}^{\dagger}\bar{\nu}^{\dagger},$ \\ 
  $X\bar{u}^{\dagger}\bar{u}\bar{\nu}\bar{\nu}\bar{\nu}\bar{\nu}\bar{\nu},
        X\bar{u}^{\dagger}\bar{u}\bar{\nu}^{\dagger}\bar{\nu}^{\dagger}
        \bar{\nu}^{\dagger}\bar{\nu}^{\dagger}\bar{\nu}^{\dagger},$ \\   
  $X\bar{d}^{\dagger}\bar{d}\bar{\nu}\bar{\nu}\bar{\nu}\bar{\nu}\bar{\nu},
        X\bar{d}^{\dagger}\bar{d}\bar{\nu}^{\dagger}\bar{\nu}^{\dagger}
        \bar{\nu}^{\dagger}\bar{\nu}^{\dagger}\bar{\nu}^{\dagger},$ \\
  $Xqq\bar{d}^{\dagger}\bar{\nu}\bar{\nu}\bar{\nu}\bar{\nu},
        Xq^{\dagger}q^{\dagger}\bar{d}\bar{\nu}^{\dagger}\bar{\nu}^{\dagger}
        \bar{\nu}^{\dagger}\bar{\nu}^{\dagger},$ \\
  $Xq\bar{u}l^{\dagger}\bar{\nu}\bar{\nu}\bar{\nu}\bar{\nu},
        Xq^{\dagger}\bar{u}^{\dagger}l\bar{\nu}^{\dagger}\bar{\nu}^{\dagger}
        \bar{\nu}^{\dagger}\bar{\nu}^{\dagger},$ \\
  $Xq^{\dagger}\bar{d}^{\dagger}l^{\dagger}
        \bar{\nu}\bar{\nu}\bar{\nu}\bar{\nu}, 
        Xq\bar{d}l\bar{\nu}^{\dagger}\bar{\nu}^{\dagger}
        \bar{\nu}^{\dagger}\bar{\nu}^{\dagger},$\\
  $X\bar{u}^{\dagger}\bar{d}^{\dagger}\bar{d}^{\dagger}
        \bar{\nu}\bar{\nu}\bar{\nu}\bar{\nu},
        X\bar{u}\bar{d}\bar{d}\bar{\nu}^{\dagger}\bar{\nu}^{\dagger}
        \bar{\nu}^{\dagger}\bar{\nu}^{\dagger},$ \\
  $X\bar{u}\bar{d}^{\dagger}\bar{e}\bar{\nu}\bar{\nu}\bar{\nu}\bar{\nu},
      X\bar{u}^{\dagger}\bar{d}\bar{e}^{\dagger}
      \bar{\nu}^{\dagger}\bar{\nu}^{\dagger}
      \bar{\nu}^{\dagger}\bar{\nu}^{\dagger},$ \\
  $Xq\bar{d}^{\dagger}\bar{d}^{\dagger}l^{\dagger}
      \bar{\nu}\bar{\nu}\bar{\nu},
      Xq^{\dagger}\bar{d}\bar{d}l
      \bar{\nu}^{\dagger}\bar{\nu}^{\dagger}\bar{\nu}^{\dagger},$ \\
  $X\bar{u}\bar{d}^{\dagger}l^{\dagger}l^{\dagger}\bar{\nu}\bar{\nu}\bar{\nu},
      X\bar{u}^{\dagger}\bar{d}ll
      \bar{\nu}^{\dagger}\bar{\nu}^{\dagger}\bar{\nu}^{\dagger},$ \\
  $X\bar{d}^{\dagger}\bar{d}^{\dagger}\bar{d}^{\dagger}\bar{e}
      \bar{\nu}\bar{\nu}\bar{\nu},
      X\bar{d}\bar{d}\bar{d}\bar{e}^{\dagger}
      \bar{\nu}^{\dagger}\bar{\nu}^{\dagger}\bar{\nu}^{\dagger},$ \\
  $X\bar{d}^{\dagger}\bar{d}^{\dagger}\bar{d}^{\dagger}l^{\dagger}
      l^{\dagger}\bar{\nu}\bar{\nu},
      X\bar{d}\bar{d}\bar{d}ll\bar{\nu}^{\dagger}\bar{\nu}^{\dagger};$ \\
  $W = Xq\bar{u}H_u\bar{\nu}\bar{\nu}\bar{\nu}\bar{\nu}\bar{\nu},\quad
      Xq\bar{d}H_d\bar{\nu}\bar{\nu}\bar{\nu}\bar{\nu}\bar{\nu},$ \\
  $XqqqH_d\bar{\nu}\bar{\nu}\bar{\nu}\bar{\nu},\quad
      Xq\bar{u}\bar{e}H_d\bar{\nu}\bar{\nu}\bar{\nu}\bar{\nu}.$
  \end{tabular} \\ \hline
\end{tabular}
\end{center}
\caption{The lowest dimensional operators through which the $X$ particle 
 decays into the MSSM particles and the antineutrinos for 
 $(p,m) = (0,1),(0,3)$ under the discrete ${\bf Z}_{10}$ symmetry.
(The number $n_X$ in parenthesis ${\bf Z}_{10}(n_X)$ denotes 
 ${\bf Z}_{10}$ charge for the $X$ particle.)
 For each case, the lifetime of the $X$ particle $\tau_X$ is 
 $10^{10}-10^{23}~{\rm years}$.
 The $X$ particle is assumed to transform as $({\bf 8}, {\bf 1})_0$
 under the standard-model gauge group.
 For $(p,m) = (0,2)$ and $(0,4)$, the $X$ particle is completely
 stable, since all MSSM particles and antineutrinos have even charges
 while the $X$ particle has an odd charge.
 We discard the cases $(p,m) = (0,6)-(0,9)$, since they are charge
 conjugations of the $(p,m) = (0,1)-(0,4)$.
 Here, we have set the cut-off scale $M_* = 1$ in the table.}
\label{lowest_operator_Dirac_A}
\end{table}
\end{document}